\pdfoutput=1

\documentclass{article}
\usepackage{natbib} %for bibtex
\usepackage{url} %for formatting urls
\usepackage{array} %for tabular fixed-width columns
\usepackage{amsthm, amsmath, graphicx, amsfonts, mathrsfs, float, enumerate, multirow, rotating}
\usepackage{longtable, subcaption}
\usepackage{array}
\usepackage{multicol,multirow}
\usepackage[labelfont=bf]{caption}

\usepackage{subcaption}
\DeclareCaptionSubType*[Alph]{table}
\DeclareCaptionLabelFormat{mystyle}{Table~\bothIfFirst{#1}{ ̃}#2}
\usepackage[a4paper]{geometry}
\newcolumntype{M}[1]{>{\centering\arraybackslash}m{#1}}
\newcolumntype{N}{@{}m{0pt}@{}}

\newcommand{\sumi}{\sum_{i=1}^I}
\newcommand{\sumj}{\sum_{j=1}^J}

\newcommand{\sui}{\sigma^2_{U,i}}
\newcommand{\su}{\sigma^2_{U}}

\newcommand{\suh}{\hat\sigma^2_{U}}
\newcommand{\sw}{\sigma^2_W}

\newcommand{\swh}{\hat\sigma^2_W}
\newcommand{\swjh}{\hat\sigma^2_{W_j}}
\renewcommand{\sc}{\sigma^2_X}
\newcommand{\sch}{\hat\sigma^2_X}

\begin{document}
\title{Improving Reliability of Subject-Level Resting-State fMRI Parcellation with Shrinkage Estimators}
\author{Amanda F. Mejia$^{\rm a}$, Mary Beth Nebel$^{\rm b}$, Haochang Shou$^{\rm g}$, Ciprian M. Crainiceanu$^{\rm a}$,\\James J. Pekar$^{\rm c,d}$, Stewart Mostofsky$^{\rm b,e,f}$, Brian Caffo$^{\rm a}$ and Martin A. Lindquist$^{\rm a}$\footnote{Corresponding author: 615 N. Wolfe Street, E3634; Baltimore, MD 21205, e-mail: mlindqui@jhsph.edu.}\\ \\ \\
\date{}
$^{\rm a}${\it{Department of Biostatistics, Johns Hopkins University, USA}}\\ 
$^{\rm b}${\it{Center for Neurodevelopmental and Imaging Research, Kennedy Krieger Institute, USA}}\\
$^{\rm c}${\it{Department of Radiology, Johns Hopkins School of Medicine, USA}}\\
$^{\rm d}${\it{F.M. Kirby Research Center for Functional Brain Imaging, Kennedy Krieger Institute, USA}}\\
$^{\rm e}${\it{Department of Neurology, Johns Hopkins School of Medicine, USA}}\\
$^{\rm f}${\it{Department of Psychiatry and Behavioral Sciences, Johns Hopkins School of Medicine, USA}}\\
$^{\rm g}${\it Department of Biostatistics and Epidemiology, University of Pennsylvania, USA}
}

\maketitle
\thispagestyle{empty}
\newpage

\pagestyle{plain}

\begin{abstract}
A recent interest in resting state functional magnetic resonance imaging (rsfMRI) lies in
subdividing the human brain into anatomically and functionally distinct regions of interest. For
example, brain parcellation is often a necessary step for defining the network nodes used in
connectivity studies. While inference has traditionally been performed on group-level data, there is a
growing interest in parcellating single subject data. However, this is difficult due to the inherent low
signal-to-noise ratio of rsfMRI data, combined with typically short scan lengths. A large number of
brain parcellation approaches employ clustering, which begins with a measure of similarity or distance
between voxels. The goal of this work is to improve the reproducibility of single-subject parcellation
using shrinkage-based estimators of such measures, allowing the noisy subject-specific estimator to
``borrow strength" in a principled manner from a larger population of subjects. We present several
empirical Bayes shrinkage estimators and outline methods for shrinkage when multiple scans are not
available for each subject. We perform shrinkage on raw inter-voxel correlation estimates and use
both raw and shrinkage estimates to produce parcellations by performing clustering on the voxels.
While we employ a standard spectral clustering approach, our proposed method is agnostic to the
choice of clustering method and can be used as a pre-processing step for any clustering algorithm.
Using two datasets -- a simulated dataset where the true parcellation is known and is subject-specific
and a test-retest dataset consisting of two 7-minute resting-state fMRI scans from 20 subjects -- we
show that parcellations produced from shrinkage correlation estimates have higher reliability and
validity than those produced from raw correlation estimates. Application to test-retest data shows
that using shrinkage estimators increases the reproducibility of subject-specific parcellations of the
motor cortex by up to 30\%.
\end{abstract}
%\noindent {\bf Key words:} 

\newpage
\setcounter{page}{1}
\section{Introduction}
There has been a long-standing interest in subdividing the human brain into anatomically and functionally distinct regions. Previously these subdivisions, or parcellations, were based primarily on mapping anatomical features from post-mortem brains (\citealt{zilles2010centenary}). More recently, the use of resting-state functional magnetic resonance imaging (rsfMRI) has provided the means for performing parcellation on living subjects using functional information (\citealt{bellec2010multi}; \citealt{craddock2011}; \citealt{yeo2011organization}).

There are several potential reasons for the increased interest in functional parcellation of the brain. First, it provides an atlas that can be used to more accurately compare inter-subject fMRI time series by incorporating functional and anatomical features into inter-subject registration approaches (\citealt{thirion2006dealing}).  Second, it allows for dimension reduction in fMRI analysis by grouping together functionally similar voxels, which not only reduces computational burdens, but also alleviates the problem of multiple comparisons and overly conservative family-wise error rate (FWER) corrections (\citealt{lindquist2014zen}). Third, the identification of functionally homogeneous regions of interest (ROIs) is necessary for defining meaningful brain network nodes (\citealt{sporns2005human}).

Many methods have been used to functionally parcellate the brain. These include, among others, fuzzy C-means (\citealt{baumgartner1997fuzzy}), independent components analysis (\citealt{mckeown1998independent}; \citealt{damoiseaux2006}; \citealt{deluca2006}), expectation maximization (\citealt{ryali2012}), hierarchical clustering (\citealt{cordes2002hierarchical}; \citealt{salvador2002simple}; \citealt{blumensath2013spatially}), spectral clustering (\citealt{craddock2011}; \citealt{nebel2012}), and K-means clustering (\citealt{kim2010}).  The goal of clustering is to group together items that are similar to each other and separate items that are dissimilar from each other.  As such, all clustering methods for brain parcellation employ a measure of distance or similarity that is used to classify voxels into coherent clusters.  There are many such metrics available, including euclidian distance, correlation, and eta-squared (\citealt{cohen2008}; \citealt{nebel2012}), and the choice of metric will have a direct effect on the result of clustering.  Moreover, these metrics are subject to error whenever the underlying data are measured with noise, and the degree of noise may have a strong effect on clustering results.  While the noise levels of rsfMRI data may be sufficiently low when the data from many subjects is combined or averaged to form a group-level parcellation, the noise in a single subject's data is substantially higher.  The primary approaches to overcome this limitation have been collecting greater amounts of rsfMRI data on a single subject (30-60 minutes rather than the standard 5-10 minutes) (\citealt{cohen2008}; \citealt{blumensath2013spatially}; \citealt{wig2013parcellating}) and utilizing constrained clustering algorithms (\citealt{craddock2011}; \citealt{basu2002semi}; \citealt{blumensath2013spatially}).  For example, \citeauthor{blumensath2013spatially} (\citeyear{blumensath2013spatially}) proposed a subject-level clustering method in which a set of stable seeds is grown into an initial parcellation that is further clustered using a hierarchical approach that enforces spatial contiguity. However, in this paper and others, reliability is assessed on a single subject with a large amount of scan time.  In contrast, most rsfMRI data are collected on much shorter intervals, making validation and replication of such approaches hard for practical purposes. Thus, the generalizability of such methods to more than one subject and their reliability on scans of shorter length are still open questions.  Moreover, accurate assessment of reliability and validity of constrained clustering methods is difficult as the ground truth is unknown, and constraints artificially inflate reliability metrics by reducing the degrees of freedom of the problem (\citealt{blumensath2013spatially}).

Our proposal is to directly improve the reliability of distance metrics by using shrinkage estimators. Advantages of this approach are that the amount of scan time required to produce reliable subject-level parcellations is greatly reduced; resulting data can be used with standard, efficiently implemented clustering algorithms; and clustering results are a product of only the data itself, and not of external constraints.   

The goal of this work is to investigate whether shrinkage-based methods can improve the reproducibility of subject-level parcellations generated using rsfMRI data. Shrinkage methods allow noisy subject-level estimators to ``borrow strength'' in a principled manner from a larger population of subjects. In the statistics literature, shrinkage estimators (\citealt{james1961estimation}; \citealt{efron1975data}) have been shown to improve the mean squared error (MSE) of many traditional estimators by shrinking the estimators towards some fixed constant value, such as the population mean. Shrinkage is implicit in Bayesian inference, penalized likelihood inference and multi-level models (\citealt{lindquist2009correlations}) and is directly related to the empirical Bayes estimators commonly used in neuroimaging (\citealt{friston2002classical}; \citealt{friston2003posterior}; \citealt{su2009modified}).  Recently, \citeauthor{shou2014shrinkage} (\citeyear{shou2014shrinkage}) applied shrinkage in the context of rsfMRI seed-based connectivity analysis and showed a nearly 30\% average improvement in intra-subject reliability of correlation estimates, with improvement of over 50\% in several subjects.  

This paper extends the work of \citeauthor{shou2014shrinkage} (\citeyear{shou2014shrinkage}) and offers a number of methodological contributions.  First, we develop shrinkage estimators for the full voxel-by-voxel distance matrix, which is required for clustering.  Second, we propose methods for constructing shrinkage estimators in the practical case where only a single scan is available for some or all subjects.  Third, we explore the utility of shrinkage estimators where the degree of shrinkage performed is subject-dependent to account for differences in intra-subject variability.  Finally, we perform clustering on shrinkage estimates and demonstrate improved test-retest reliability of the resulting subject-level parcellations. 

To illustrate the feasibility of our proposed approach, we focus on one simple unsupervised learning technique, namely normalized spectral clustering. We generate simulated data, where the true parcellation is known and is allowed to vary across subjects.  In addition, we apply the method to real test-retest resting-state fMRI data from 20 subjects and show that we can increase the reliability of single-subject parcellations of the motor cortex by up to 30\%.

\section{Methods}

In this section, we discuss shrinkage methods and illustrate how they can be used for single-subject parcellation.  We begin by detailing our shrinkage model and methods for estimating the relevant parameters.  We perform shrinkage on measures of functional connectivity and obtain parcellations from these measures using two sets of data, which are described below.  The first is a simulation, for which the ground truth parcellation is known. The second is a test-retest dataset of resting-state fMRI scans, for which we use part of each subject's data as a proxy for the unknown ground truth parcellation.  

\subsection{Shrinkage Methods}
\label{sec:shrinkageMethods}

\subsubsection{Shrinkage Model}
\label{sec:shrinkageModel}

The quantity of interest for each subject $i$, $i=1,\dots,I$, is the true $V \times V$ functional connectivity matrix $C_i$, where $V$ is the number of voxels in a pre-defined region of interest (ROI) of the brain, which we wish to parcellate.  Our goal is to use information from the other $I-1$ subjects to provide stable estimates of the subject-specific value $C_i$; the idea is to find compromise estimators between the unbiased but highly variable raw subject-specific connectivity estimators and the biased, but much smoother, mean connectivity estimators.

More precisely, let $v$ and $v'$ be two distinct voxels in a particular ROI, let $X_i(v,v')$ be the true quantity of interest for subject $i$ (e.g. pairwise correlation), and let $W_{ij}(v,v')$ be the observed value of $X_i(v,v')$ obtained from session $j$.  The classical measurement error model (\citealt{carroll2006measurement}) is
$$
W_{ij}(v,v') = X_i(v,v') + U_{ij}(v,v'),
$$ 
where $U_{ij}(v,v')$ is subject-level measurement error for subject $i$ during session $j$ at voxel-pair $(v,v')$.  We assume that $X_i(v,v')$ and $U_{ij}(v,v')$ are independent for all $i$ and $j$.  We further assume that the $X_i(v,v')$, $i=1,\dots,I$, are independently drawn from a population distribution with between-subject variance $\sc(v,v')$, i.e.
$$
X_i(v,v')\sim N\left\{\mu_X(v,v'),\sc(v,v')\right\}. 
$$ 
Finally, we assume that for each subject $i$, the $U_{ij}(v,v')$ are independently and identically distributed for all $j$ and
$$
U_{ij}(v,v')\sim N\left\{0,\sui(v,v')\right\}. 
$$ 
Returning our attention to the quantity of interest, the shrinkage estimator of $X_i(v,v')$ using session $j$ is given by 
$$
\tilde{W}_{ij}(v,v') = \lambda_i(v,v') * \bar{W}_j(v,v') + \{1-\lambda_i(v,v')\} * W_{ij}(v,v'),
$$
where $\bar{W}_j(v,v') = \frac{1}{I}\sum_{i=1}^I W_{ij}(v,v')$, and the shrinkage parameter $\lambda_i(v,v')$ represents the relationship between within-subject variance $\sui(v,v')$ and between-subject variance $\sc(v,v')$:
$$
\lambda_i(v,v')
= \frac{\sui(v,v')}{\sc(v,v')+\sui(v,v')}.
$$  
Here $\lambda_i(v,v')$ ranges between 0 and 1 and represents the relative weight given to the group mean $\bar{W}_j(v,v')$ compared to the raw subject-level estimate $W_{ij}(v,v')$.  As the within-subject variance $\sui(v,v')$ increases, the subject-level information is less reliable, the shrinkage parameter increases, and the shrinkage estimate is more weighted towards the group mean.  As between-subject variance $\sc(v,v')$ increases, the group mean becomes less representative of the true subject-level values, so shrinkage is less beneficial, $\lambda_i(v,v')$ decreases, and the shrinkage estimate is more weighted towards the subject-level observation.  $\lambda_i(v,v')$ is estimated directly from the data and is designed to achieve the optimal balance between the raw subject-level estimate and the group mean.

We employ the Fisher-transformed correlation coefficient as our measure of functional connectivity, which fulfills the model assumptions of Normality and independence of $X_i(v,v')$ and $U_{ij}(v,v')$ (\citealt{shou2014shrinkage}).  Given a correlation estimate $r$, the Fisher-transformed estimate $z(r)$ is given by the transformation $r \to \frac{1}{2}\log\left(\frac{1+r}{1-r}\right)$, and is approximately Normally distributed with variance $(T-3)^{-1}$, where $T$ is the number of time points in the scan.  By contrast, the sampling variance of an untransformed correlation coefficient decreases as the true correlation increases, which violates the signal-noise independence assumption.  After shrinkage is performed, we then apply the inverse transformation $z \to \frac{exp(2z)-1}{exp(2z)+1}$ to obtain an estimate of correlation for the purposes of parcellation.  However, for completeness we also evaluate the benefits of applying shrinkage directly to the untransformed correlation estimates.

Note that the within-subject variance $\sui(v,v')$ is allowed to vary across subjects $i$.  This allowance stems from the observation that within-subject variance comes from multiple sources, including sampling variability and session-to-session variability.  Sampling variability reflects the error of an estimate (e.g. correlation estimate) around the value it is estimating and is directly related to the number of time points used to compute the estimate.  For example, as described above, a Fisher-transformed correlation estimate has asymptotic sampling variance $\frac{1}{T-3}$.  It follows that as the number of time points increases to infinity, this source of variability will decrease to zero.  While sampling variability may be roughly equal across subjects with equal scan lengths, session-to-session variability may vary across subjects.  This type of variability reflects differences in a subject's true functional connectivity across multiple scanning sessions due to variations in brain behavior.  Moreover, session-to-session variability may dominate sampling variability.  In our sample, for example, we find that within-subject variance of the Fisher-transformed correlation matrices tends to be around five times larger than the theoretical sampling variance.  We therefore allow $\sui(v,v')$ to differ across subjects.  However, as there are drawbacks to estimating $\sui(v,v')$ completely separately for each subject, we propose several other methods of estimating within-subject (``noise'') variance.  In total, we propose four methods, which are discussed in detail below.

\subsubsection{Variance Component Estimation}
\label{sec:Variance}

Henceforth, we will use the terms \textit{within-subject} and \textit{noise} variance interchangeably, and we will use the terms \textit{between-subject} and \textit{signal} variance interchangeably.

\textbf{Noise Variance Estimation}

To estimate the noise variance, it is ideal to have access to multiple scanning sessions for each subject (``test-retest data'').  However, in many cases only a single scan is available for each subject.  For these situations, we propose two approaches.  First, create a \textit{pseudo}-test-retest dataset by dividing each subject's single scan into two sub-scans, each containing half of the original time points.  Second, estimate a global measure of within-subject variance for all voxel-pairs and subjects using an external test-retest dataset, a subset of subjects for which multiple sessions are available, or through psuedo-test-retest data combined with extrapolation.  This will be discussed in more detail below.

We now describe four noise variance estimators, which we denote common (C), individual (I), scaled (S), and global (G).  The common estimator assumes that $\sui(v,v')\equiv\su(v,v')$ is the same across all subjects, while allowing variation across voxel-pairs.  The individual and scaled estimators allow $\sui(v,v')$ to vary across subjects and voxel-pairs.  The individual estimator estimates $\sui(v,v')$ separately for each subject, while the scaled estimator starts with the common noise variance estimator and adjusts it by a subject-specific factor to produce a different noise variance estimate for each subject.  The global estimator assumes that $\sui(v,v')=\su$ is the same across all subjects and voxel-pairs. This estimator is primarily intended for the case when limited or no test-retest data is available for the dataset of interest.  All four noise variance estimators can be computed using true test-retest data or pseudo-test-retest data created from a single scan.

\textit{Common Noise Variance}\\
Letting $D_i(v,v')=W_{i2}(v,v')-W_{i1}(v,v')$, the common noise variance can be estimated as (\citealt{carroll2006measurement}, \citealt{shou2014shrinkage})
$$
\hat\sigma_{U,i}^{2(C)}(v,v')\equiv\hat\sigma_U^{2(C)}(v,v'):=\frac{1}{2}Var_i\left\{D_i(v,v')\right\}=\frac{1}{2(I-1)}\sumi\left\{D_i(v,v')-\bar{D}(v,v')\right\}^2,
$$
where $\bar{D}(v,v')=\frac{1}{I}\sumi D_i(v,v')$.  To see this, note that 
\begin{align*}
Var_i\left\{D_i(v,v')\right\}&=Var_i\left\{W_{i2}(v,v')-W_{i1}(v,v')\right\} \\
&= 2Var_i\left\{U_{ij}(v,v')\right\},
\end{align*}
so $Var_i\left\{U_{ij}(v,v')\right\}=\frac{1}{2}Var_i\left\{D_i(v,v')\right\}$.  

\textit{Individual Noise Variance}\\
Given two estimates $W_{ij}(v,v')$, $j=1,2$, of the term $X_i(v,v')$, the individual noise variance can be estimated as follows:
\begin{align*}
\hat\sigma_{U,i}^{2(I)}(v,v') 
&:=\frac{1}{J-1}\sum_{j=1}^J \left\{U_{ij}(v,v')-\bar{U_i}(v,v')\right\}^2 \\
&=\left\{U_{i1}(v,v')-\bar{U_i}(v,v')\right\}^2 + \left\{U_{i2}(v,v')-\bar{U_i}(v,v')\right\}^2 \\
&=\frac{1}{2}\left\{U_{i2}(v,v')-U_{i1}(v,v')\right\}^2 \\
&=\frac{1}{2}\big\{\big[W_{i2}(v,v')-X_i(v,v')\big]-\big[W_{i1}(v,v')-X_i(v,v')\big]\big\}^2 \\
&=\frac{1}{2}\left\{W_{i2}(v,v')-W_{i1}(v,v')\right\}^2
\end{align*}

\textit{Scaled Noise Variance}\\
Given the common noise variance estimate $\hat\sigma_U^{2(C)}(v,v')$, we use a subject-specific scaling factor $\gamma_i$ to obtain the scaled noise variance estimate
$$
\hat\sigma_{U,i}^{2(S)}(v,v')=\gamma_i\times\hat\sigma_U^{2(C)}(v,v').
$$
The scaling factor $\gamma_i$ is equal to the test-retest MSE of subject $i$ relative to the average test-retest MSE over all subjects:
$$
\gamma_i=\bar{D^2}_i/\bar{\bar{D^2}},
$$
where $\bar{D^2}_i=\frac{2}{V(V-1)}\sum_{v>v'}D_i^2(v,v')$ and $\bar{\bar{D}}=\frac{1}{I}\sum_{i=1}^I \bar{D^2}_i$.  To see that this provides a reasonable estimate of an individual subject's noise variance, notice that
$$
\hat\sigma_{U,i}^{2(I)}(v,v')=\frac{1}{2}D_i^2(v,v')
$$
and
$$
\hat\sigma_U^{2(C)}(v,v') \approx 
\frac{1}{I}\sumi \hat\sigma_{U,i}^{2(I)}(v,v')=\frac{1}{2I}\sumi D_i^2(v,v')=\frac{1}{2}\bar{D^2}(v,v').
$$
The approximate equality above follows from the fact that the two terms have the same expected value (Appendix).  Therefore, 
$$
\frac{\hat\sigma_{U,i}^{2(I)}(v,v')}{\hat\sigma_U^{2(C)}(v,v')}\approx \frac{D_i^2(v,v')}{\bar{D^2}(v,v')},
$$
so across all voxel-pairs $(v,v')$, the ratio of the the individual noise variance to the common noise variance is approximately equal to $\gamma_i$.  The benefit of the scaled noise variance, compared with the individual noise variance, is that it is based upon a more stable estimate of the noise variance, $\hat\sigma_U^{2(C)}(v,v')$ and requires the estimation of much fewer terms.

\textit{Global Noise Variance}\\
We estimate the global noise variance as the mean value of the common noise variance over all $V(V-1)/2$ unique voxel-pairs:
$$
\hat\sigma_{U,i}^{2(G)}(v,v')\equiv\hat\sigma_U^{2(G)}:=\frac{2}{V(V-1)}\sum_{v>v'}\hat\sigma_U^{2(C)}(v,v').
$$
If pseudo-test-retest data is used to compute $\hat\sigma_U^{2(G)}$, the noise variance will be overestimated due to shorter scan length and should be adjusted.  Let $\sigma_U^2(t)$ be the expected global noise variance for a scan of length $t$, and let $\hat\sigma_U^2(t)$ be an estimate of $\sigma_U^2(t)$.  When we use pseudo-test-retest data by splitting a scan of length $T$, we obtain an estimate of $\sigma_U^2(\tfrac{T}{2})$.  To obtain an estimate of $\sigma_U^2(T)$, let
$$
\theta(T)=\frac{\sigma_U^2(T)}{\sigma_U^2(\tfrac{T}{2})}.
$$
Then $\sigma_U^2(T)$ can be estimated as
\begin{equation} \label{eq:use_theta}
\hat\sigma_U^2(T)=\hat\theta(T)\times\hat\sigma_U^2(\tfrac{T}{2}).
\end{equation}
The adjustment factor $\theta(T)$ can be estimated if multiple scans are available for a subset of subjects.  However, we also provide an estimate of $\theta(t)$ as follows.  Using the test-retest fMRI dataset described below, for $t=\{1,1.5,2,2.5,3,3.5,4,5,6,7\}$ minutes, we resample scans of length $t$ within each scanning session and estimate $\sigma_U^2(t)$ for each resampled dataset.  We then compute the average over all resampled datasets to obtain $\hat\sigma_U^2(t)$.  We compute $\hat\theta(t)=\hat\sigma_U^2(t)/\hat\sigma_U^2(\tfrac{t}{2})$ for $t=\{2,3,4,5,6,7\}$.  This gives a curve estimating the relationship between $T$ and $\theta(T)$ for $T=\{2,3,4,5,6,7\}$.  To extrapolate to other scan lengths, we fit a regression curve relating log scan length to $\theta(t)$:
\begin{equation} \label{eq:estimate_theta}
\theta(t)=\beta_0+\beta_1\times \log(t).
\end{equation}
Using our coefficient estimates of $\beta_0$ and $\beta_1$, one can estimate the appropriate scaling factor $\theta(T)$ for scans of length $T$ by simply plugging in their values into equation \ref{eq:estimate_theta}.  One can then use this scaling factor to adjust the global noise variance estimate obtained from pseudo-test-retest data using equation \ref{eq:use_theta} and hence obtain an appropriate noise variance estimate.

\textbf{Signal Variance Estimation}

The between-subject or \textit{signal} variance $\sc(v,v')$ is equal to the difference between the total variance and noise variance.  While noise variance may vary across subjects, signal variance is a population parameter.  Therefore, even if we choose to estimate the noise variance individually, we use the common or global noise variance estimator to obtain an estimate of the signal variance.  

The total variance $\sw(v,v')$ at voxel-pair $(v,v')$ is estimated as (\citealt{carroll2006measurement})
$$
\swh(v,v'):=\frac{1}{J}\sumj \swjh(v,v') = \frac{1}{J(I-1)}\sumj\sumi\left\{W_{ij}(v,v')-\bar{W}_j(v,v')\right\}^2.
$$

We can then simply estimate the signal variance as
$$
\sch(v,v') = \swh(v,v') - \suh(v,v'),
$$

\textbf{Shrinkage Parameter Estimation}

We obtain four estimators for the shrinkage parameter $\lambda_i(v,v')$ corresponding to the four noise variance estimators:
$$
\lambda_i^{(M)}(v,v')= \frac{\sigma_{U,i}^{2(M)}(v,v')}{\sigma_X^2(v,v')+\sigma_{U,i}^{2(M)}(v,v')},
$$
where $M\in\{I,S,C,G\}$.

\subsection{Subject-level Parcellations}
\label{sec:Parcellation}

Shrinkage estimates of correlation were obtained by applying the inverse-Fisher transformation to the shrinkage estimate of the Fisher-transformed correlation.  That is, we first Fisher-transform the raw estimate, perform shrinkage, then apply the inverse Fisher transformation.  We then generated subject-level parcellations by performing spectral clustering as described by \citeauthor{ng2001spectral} (\citeyear{ng2001spectral}), using the raw and shrinkage correlation estimates as a metric of similarity.  We chose to look for five clusters based on previously published findings (using the same test-retest resting state data set) that this was the optimal number of functional partitions for the precentral gyrus in terms of test-retest reliability (\citealt{nebel2012}). 

\subsection{Performance of shrinkage methods}
\label{sec:Performance}

\subsubsection{Reliability of functional connectivity measures}

We define reliability of a functional connectivity measure (e.g. correlation) as the MSE between the estimated measure $\hat{C}_i$ and the truth $C_i$.  We assess the performance of a shrinkage estimate as the percent decrease in MSE of the shrinkage estimate relative to the MSE of the raw estimate. 

\subsubsection{Reliability of parcellations}

We define reliability of a parcellation estimate as the Dice similarity coefficient compared with the true parcellation.  Let $\hat{A}_i$ be the adjacency matrix obtained from clustering, where $\hat{A}_i(v,v')=1$ if $v$ and $v'$ are assigned to the same parcel for subject $i$ and 0 otherwise.  Let $A_i$ be the adjacency matrix corresponding to the true parcellation of subject $i$.  Dice's coefficient of similarity between the estimated and true parcellations is defined as
$$
S(A_i,\hat{A}_i) = \frac{2|A_i\cap \hat{A}_i|}{|A_i|+|\hat{A}_i|}.
$$

We assess the performance of a parcellation obtained from a shrinkage estimate of functional connectivity as the percent increase in Dice coefficient relative to the parcellation obtained from the corresponding raw estimate.  

For the simulation described below, the true connectivity matrix and parcellation are known quantities, so we can compute exactly the MSE and Dice coefficient of the raw and shrinkage estimates.  For our fMRI dataset, however, the true connectivity matrix and parcellation are unknown.  We get around this by reserving part of each subject's data as a proxy for the truth, which we call the test set.  We compute raw and shrinkage estimates for the remaining data and compare both estimates to the raw estimate from the test set.  

\subsection{Data}

\subsubsection{Simulated Data}

\begin{figure}
\begin{subfigure}[b]{1\textwidth}
\centering
\includegraphics[scale=0.5, trim=0 10mm 5mm 0, clip]{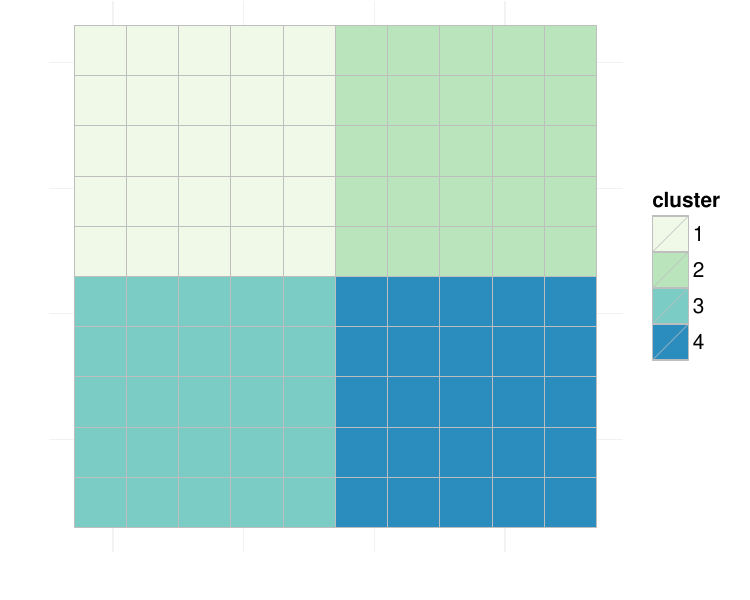}\\
\caption{Group parcellation}
\label{ClusterImg_group}
\end{subfigure}
\begin{subfigure}[b]{1\textwidth}
\centering
\includegraphics[scale=0.5, trim=0 10mm 20mm 0, clip]{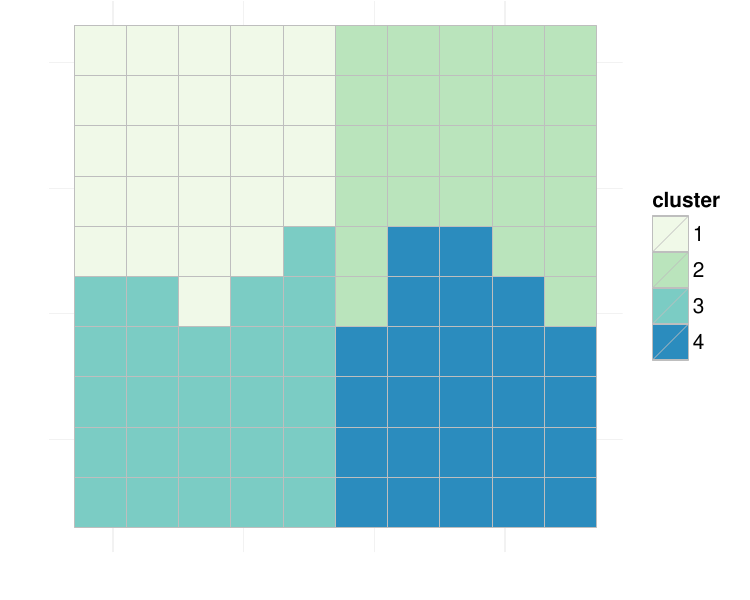}
\includegraphics[scale=0.5, trim=0 10mm 20mm 0, clip]{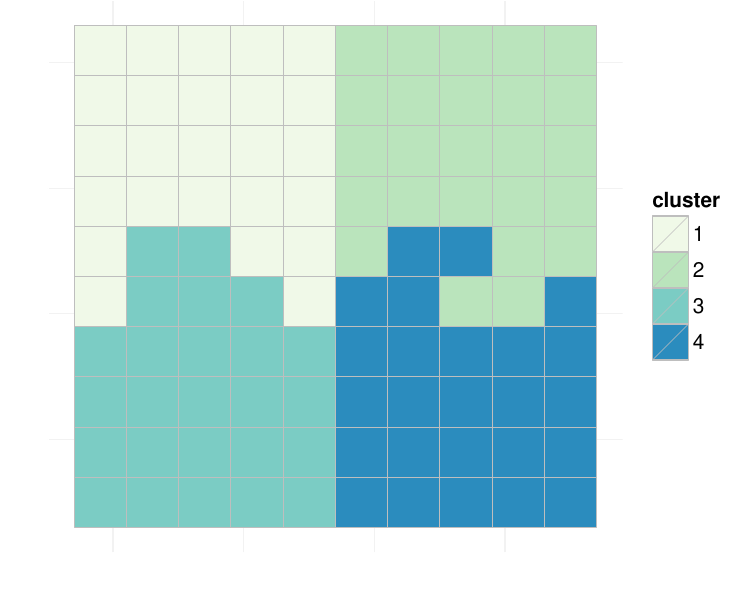}
\caption{Subject parcellations}
\label{ClusterImg_subj}
\end{subfigure}
\caption{Simulated signal image consisting of four 25-voxel clusters for the group and two subjects.}
\label{ClusterImg}
\end{figure}

We simulated a 10-by-10 voxel parcellation consisting of four clusters, each cluster corresponding to one quadrant of the image at the group level (Figure \ref{ClusterImg_group}).  Each subject-level parcellation was generated by randomly permuting cluster labels along the borders of clusters 1 and 3 and clusters 2 and 4 (Figure \ref{ClusterImg_subj}).

The true correlation matrices within each cluster were assumed to follow an \textit{exchangeable} structure, meaning that each pair of voxels within cluster $k$ of subject $i$ at session $j$ has the same pairwise correlation $\rho_{ijk}$.  We further assumed that each subject $i$ has a fixed within-cluster correlation value $\rho_{ijk}\equiv\rho_i$ across all sessions $j$ and clusters $k$.  Let $\rho$ represent the population average within-cluster correlation and $z(\rho)$ represent the Fisher-transformation of $\rho$.  Random variation among subjects $i=1,\dots,I$ was introduced by adding Gaussian noise to $z(\rho)$, then applying the inverse Fisher-transformation, $z^{-1}(\cdot)$:
$$
\rho_i = z^{-1}\big(z(\rho) + u_i\big),\ u_i\sim N(0,\sigma^2_X)
$$
As negative within-cluster correlations do not make sense under an exchangeable correlation structure, any negative correlations generated through this process were resampled until all $\rho_i$ were positive. Between-cluster correlations were assumed to equal zero.  The true correlation matrix $C_i^*$ for each subject $i$ was therefore constructed as a block diagonal matrix with the four diagonal blocks corresponding to the within-cluster correlation matrices for subject $i$, and the off-diagonal blocks set to zero.

For each subject $i=1,\dots,I$ and session $j=1,2$, a time series of length $T$ was generated for each voxel in the cluster.  Each time point was drawn from a multivariate Normal distribution with mean zero and covariance matrix $\Sigma_i\equiv C_i$.  As correlations are agnostic to within-voxel variance and temporal correlation, these were not considered.  The observed voxel time series were combined to form a 3D image (2D x time) for each subject $i$ and session $j$.  Observed correlation matrices $\hat{C}_{ij}$ were computed from the 3D images.

We varied the following simulation parameters: number of subjects ($I$), length of time series ($T$), population average within-cluster correlation ($\rho$), and between-subject or signal variance ($\sigma_X^2$). The parameter values tested are given in Table \ref{tab:ParamValues}.  Parameter values were changed one at a time, while all other parameters were fixed at a default value.  The default value of each parameter is also shown in Table \ref{tab:ParamValues}.

\begin{table}
\centering
\begin{tabular}{lll}
  \hline
Parameter & Values & Default Value  \\ 
  \hline
number of subjects ($I$) & $(10, 20, 30, 100)$ & 20  \\ 
length of time series ($T$) & $(100, 200, 300, 1000)$ & 200  \\ 
population average within-cluster correlation ($\rho$) & $(0.01,0.05,0.1)$ & 0.05  \\ 
between-subject variance ($\sigma^2_X$) & $(0.01,0.02,0.03,0.04,0.05)$ & 0.02 \\
\hline
\end{tabular}
\caption{Simulation parameters varied (one at a time) and the values they are varied over.  The default value is the value at which each parameter is fixed while the other parameters are varied.}
\label{tab:ParamValues}
\end{table}

We simulated 1000 datasets for each of the 13 unique designs defined by the parameter values in Table \ref{tab:ParamValues}.  For each dataset, we computed the observed correlation matrix $\hat{C}_{i1}$ from each subject's first session.  We performed shrinkage on these matrices using each noise variance estimation method $M\in\{I,S,C,G\}$, described in Section \ref{sec:Variance}.  For each method $M$, we computed the noise variance assuming that two sessions $j=1,2$ were available for each subject and again assuming that only one session $j=1$ was available for each subject. Let $\tilde{C}_i^{M,\ell}$ be the shrinkage estimate obtained by shrinking estimate $\hat{C}_{i1}$ using noise variance estimation method $M$ and assuming availability of $\ell=1,2$ scans for each subject.  We performed clustering as described in Section \ref{sec:Parcellation} on the raw correlation matrices $\hat{C}_{i1}$, $i=1,\dots,I$ and each shrinkage correlation matrix $\tilde{C}_i^{M,\ell}$, $M\in\{I,S,C,G\}$, $\ell=1,2$ and $i=1,\dots,I$.

\textbf{Analysis S1: Performance of shrinkage estimates and parcellations}
\label{sec:AnalysisS1}

Using the default design specified in Table \ref{tab:ParamValues}, we computed the degree of shrinkage, performance of the raw and shrinkage correlation matrices, and performance of parcellations obtained using shrinkage estimates, as described in Section \ref{sec:Performance}.  The degree of shrinkage is defined as the average value of $\lambda_i(v,v')$ over all voxel-pairs. 

\textbf{Analysis S2: Sensitivity to simulation parameters}

For each alternative design specified in Table \ref{tab:ParamValues}, we  computed the degree of shrinkage and performance of correlation estimates and parcellations to understand how each parameter affects the degree of shrinkage towards the group mean and the impact of the shrinkage procedure on the reliability of similarity metrics and parcellations. 

\subsubsection{Real fMRI Data}

\begin{figure}
\begin{subfigure}[b]{0.5\textwidth}
\centering
\includegraphics[scale=0.3]{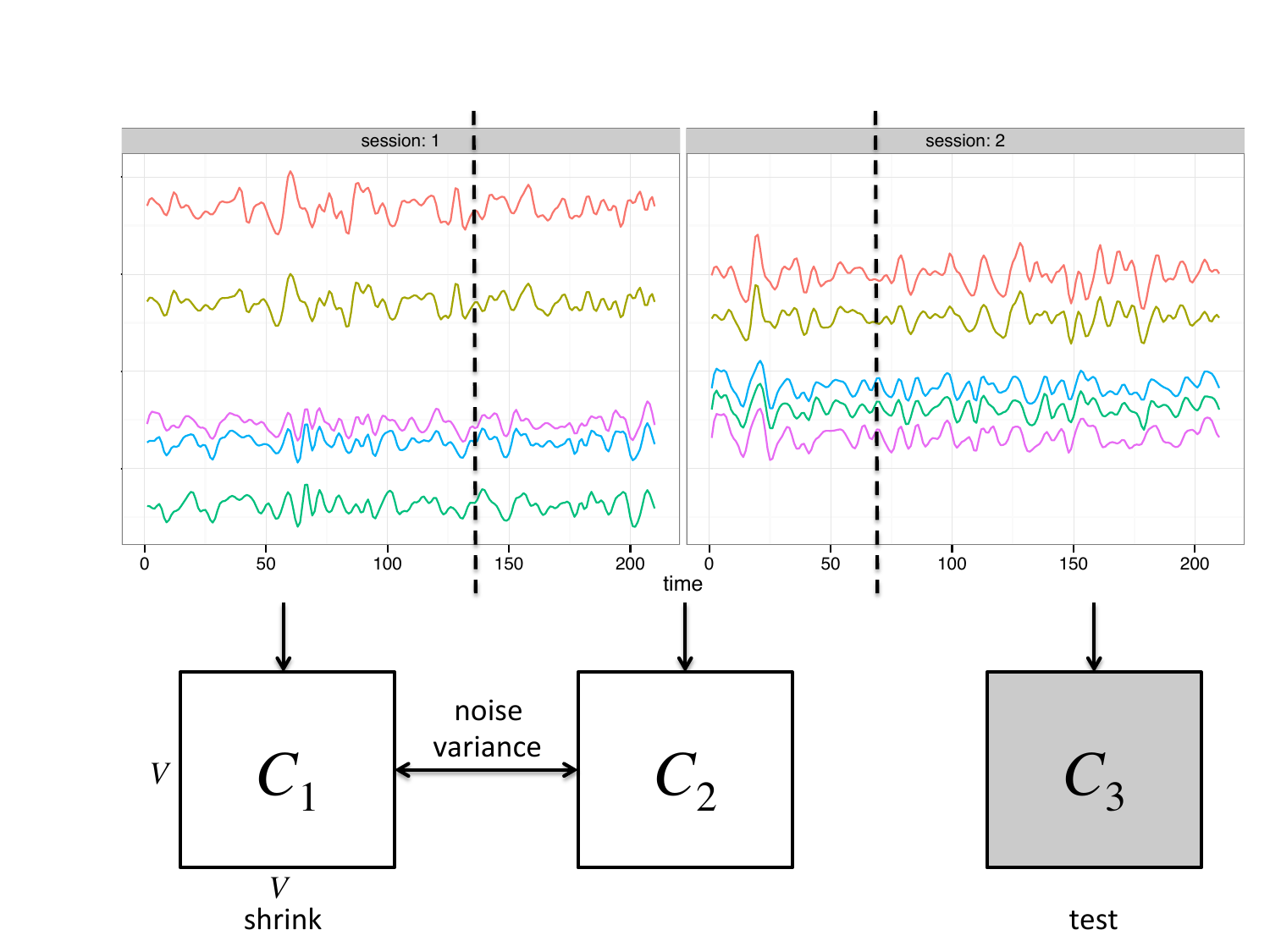}\\
\caption{Data setup for methods using test-retest data}
\label{Data_test-retest}
\end{subfigure}
\begin{subfigure}[b]{0.5\textwidth}
\centering
\includegraphics[scale=0.3]{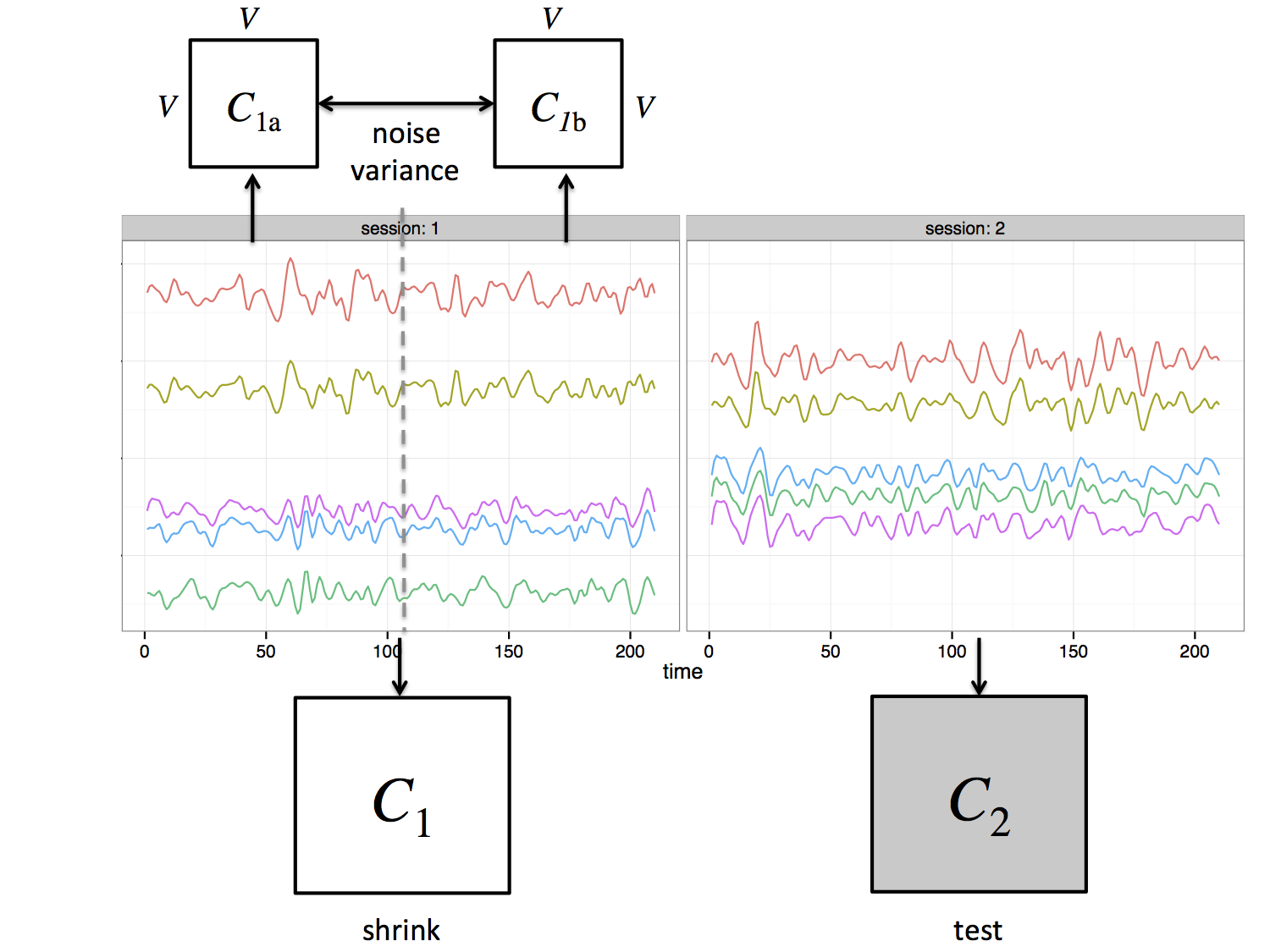}
\caption{Data setup for methods using single-scan data}
\label{Data_single-scan}
\end{subfigure}
\caption{Data setup to perform shrinkage and evaluate the performance using both test-retest data (a) and single-scan data (b).  Resting-state fMRI time series for 5 voxels are shown.}
\label{Data_setup}
\end{figure} 

We use data from the publicly available Multi-Modal MRI Reproducibility Resource (\url{http://www.nitrc.org/projects/multimodal}).  Image acquisition parameters are described in detail elsewhere (\citealt{landman2011multi}). In short, a high resolution T1-weighted MPRAGE and two 7-minute resting state scans were acquired from 21 healthy adult volunteers.  Both resting state scans were acquired on the same day, and in between the two scans the subject exited the scanner. 

The anatomical scan was registered to the first functional volume and normalized to Montreal Neuological Institute (MNI) space using SPM8's unified segmentation/normalization procedure. Resting state data were adjusted for slice time acquisition, and rigid body realignment estimates were calculated with respect to the first functional volume to account for participant motion. The non-linear spatial transformation estimated during the unified segmentation/normalization was then applied to the functional data along with the estimated rigid body realignment parameters and resulted in 2-mm isotropic voxels. Each resting state scan was then temporally detrended on a voxelwise basis. An aCompCor strategy was used to estimate spatially coherent noise components, as this method has been shown to effectively attenuate physiological noise (\citealt{behzadi2007component}) as well as motion artifacts (\citealt{muschelli2014reduction}). The aCompCor noise components were regressed from the resting state data along with linearly detrended versions of the rigid body realignment parameters and their first derivatives (computed by backward differences). Functional data were then spatially smoothed (6-mm FWHM Gaussian kernel) and temporally filtered using a .01-.1 Hz pass band. Data from one participant were excluded from analysis due to a misalignment of the first and second resting-state scans.

Our region of interest (ROI) for this experiment is the precentral gyrus (M1), a key component of the motor control network and a region whose gross functional organization has long been recognized (\citealt{penfield1937somatic}). The precentral gyrus ROI was selected from the ``Type II Eve Atlas'' (\citealt{oishi2009atlas}) and contained $V=7396$ voxels after being transformed to MNI space.

To evaluate performance of shrinkage using test-retest data to estimate the variance components, we split each subject's data into three parts as illustrated in Figure \ref{Data_test-retest}.  As a total of 420 images (14 minutes) were collected over the two sessions, each of the three parts consisted of 140 images (4 minutes and 40 seconds).  For the middle third, each session was demeaned before concatenating the time series.  The first two parts were used to compute the variance components.  We performed shrinkage on the first part and reserved the third part as the test set.  

To evaluate performance of shrinkage using only a single scan from each subject, we split the first session in half to create a pseudo-test-retest dataset, which we use to estimate the noise variance.  We performed shrinkage on the first session and reserved the second session as the test set (Figure \ref{Data_single-scan}).  The psuedo-test-retest dataset was used to compute the common, individual and scaled noise variance estimates; the global noise variance was computed using both full sessions.

For both setups, we first computed the $V$-by-$V$ observed correlation matrices $\hat{C}_{ij}$ for each subject $i$ and session $j=1,2$ or part $j=1,2,3$.  The estimates produced using the first session or part, $\hat{C}_{i1}$, were treated as the ``raw'' estimates.  We then applied the Fisher-transformation to obtain $\hat{Z}_{ij}$ for all $i$ and $j$.  We computed the variance components and shrinkage parameter $\lambda_i^{(M)}(v,v')$, $M\in\{I,S,C,G\}$, and performed shrinkage using $\lambda_i^{(M)}(v,v')$ on the $\hat{Z}_{i1}$.  We then applied the inverse Fisher transform to obtain shrinkage estimates $\tilde{C}_{i1}$.  For completeness, we also performed the same procedure directly on the $\hat{C}_{i1}$ without Fisher-transforming to obtain shrinkage estimates $\tilde{\tilde{C}}_{i1}$.

\textbf{Analysis R1: Performance of shrinkage estimates}

The performance of the shrinkage correlation estimates computed from session or part 1 was assessed as described in Section \ref{sec:Performance}, using the raw estimates from the test set as a proxy for the unknown ground truth. 

\textbf{Analysis R2: Performance of parcellations}

Subject-level parcellations were generated using both raw and shrinkage correlation estimates, as described in Section \ref{sec:Parcellation}.  We used the estimates obtained by applying shrinkage directly to the correlation estimates, as these were shown to have better performance than shrinkage estimates obtained through the Fisher-transformed correlation estimates.  The performance of the parcellations generated using shrinkage estimates from session or part 1, relative to the parcellations generated using the corresponding raw estimates, was assessed as described in Section \ref{sec:Performance}, using the parcellations generated from the raw estimates from the test set as a proxy for the ground truth.

\section{Results}

\subsection{Simulation Results}

\begin{figure}
\begin{subfigure}[b]{1\textwidth}
\centering
\includegraphics[scale=0.5]{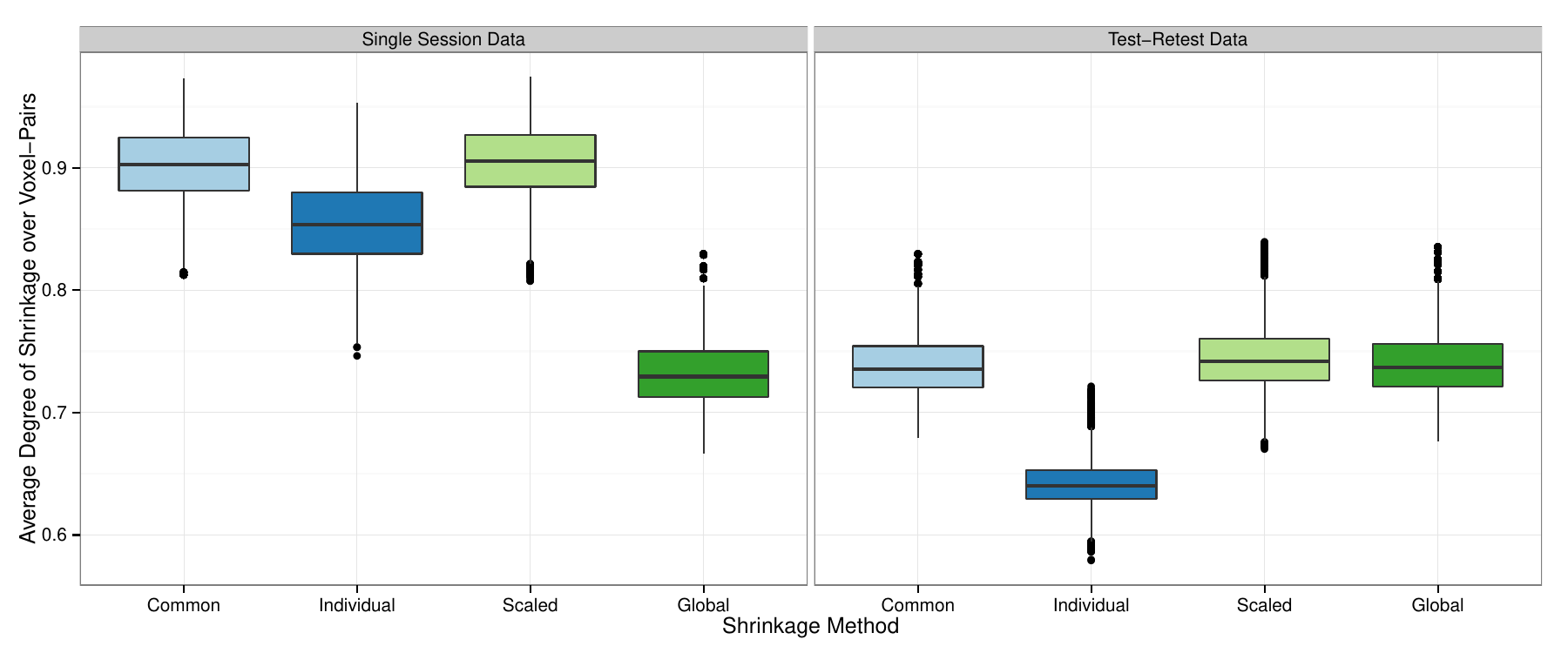}
\caption{Degree of shrinkage}
\label{fig:simulation1a}
\end{subfigure}
\begin{subfigure}[b]{1\textwidth}
\centering
\includegraphics[scale=0.5]{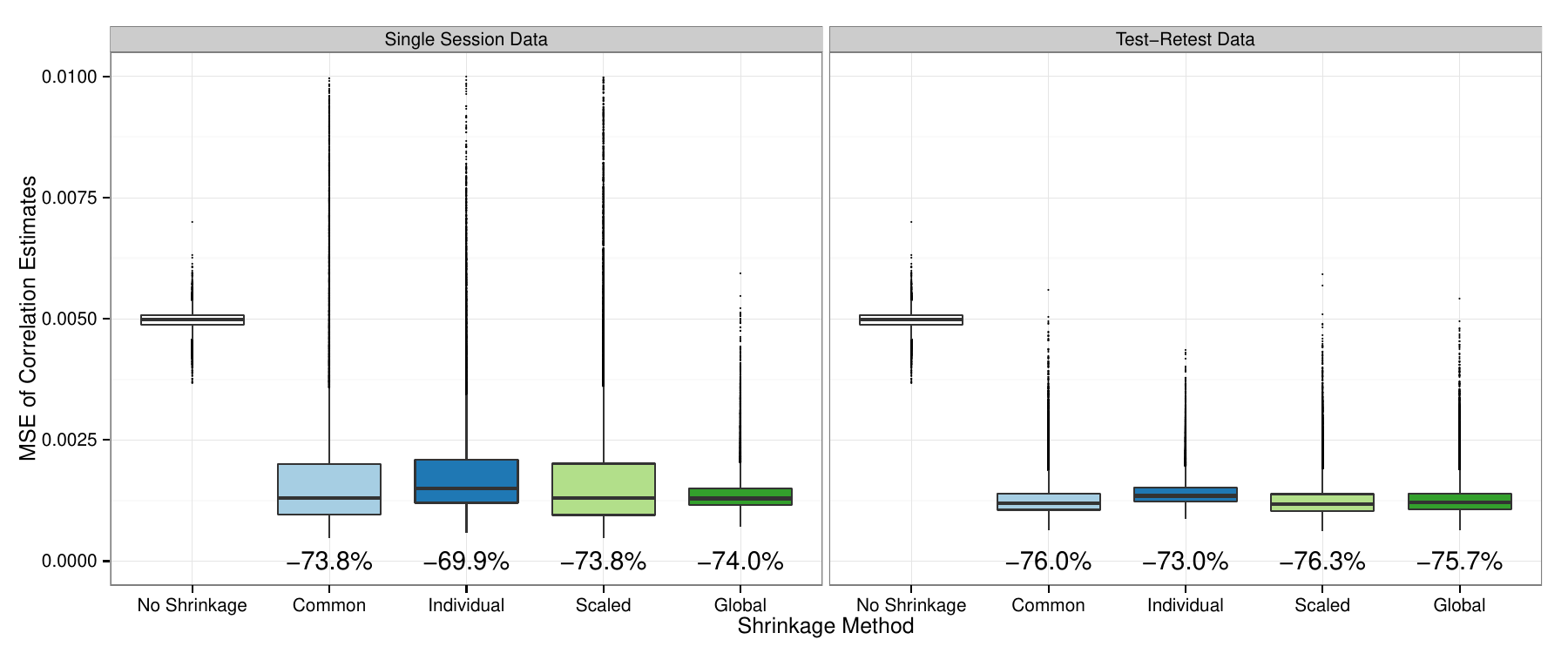}
\caption{MSE of Correlation Estimates}
\label{fig:simulation1b}
\end{subfigure}
\begin{subfigure}[b]{1\textwidth}
\centering
\includegraphics[scale=0.5]{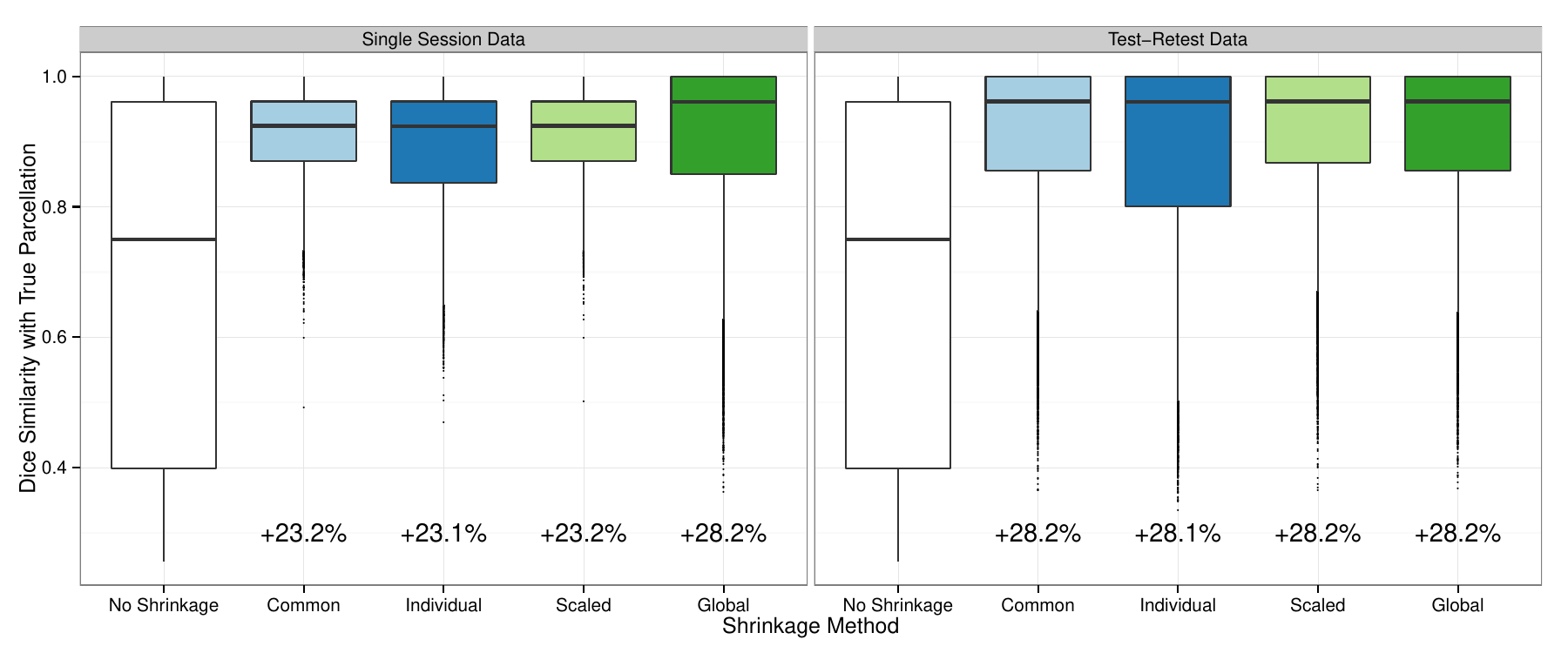}
\caption{Dice Similarity of Parcellations}
\label{fig:simulation1c}
\end{subfigure}
\caption{The degree of shrinkage (a), MSE of correlation estimates (b), and Dice coefficient of similarity of parcellations with the true parcellations (c) under the default simulation parameters of $I=20$, $T=200$, $\rho=0.05$, and $\sigma^2_X=0.02$, using either single session data (left) or test-retest data (right).  The percent decrease in the median MSE of the shrinkage estimates, compared with the MSE of the raw estimates, is reported below each boxplot in (b); the percent increase in the Dice coefficient of the shrinkage-based parcellations, compared with the Dice coefficients of the raw parcellations, is reported below each boxplot in (c).}
\label{fig:simulation1}
\end{figure}

\textbf{Analysis S1: Performance of shrinkage estimates and parcellations}

Figure \ref{fig:simulation1} shows the degree of shrinkage, MSE of raw and shrinkage correlation estimates, and Dice similarity of parcellations with the true parcellations, under the default simulation settings ($I=20$, $T=200$, $\rho=0.05$, $\sigma_X^2=0.02$).  Results are shown by shrinkage method and the type of dataset used, single session (left) or test-retest data (right), and each boxplot shows the distribution of values over all subjects and simulation iterations.  In Figure \ref{fig:simulation1a}, the degree of shrinkage for a given subject and simulation iteration was computed as the average value of the shrinkage parameter $\lambda_i(v,v')$ over all voxel-pairs $(v,v')$.  The degree of shrinkage was sensitive to the noise variance method employed and the type of dataset (single session or test-retest) used to perform shrinkage.  Using a single session to perform shrinkage, the median degree of shrinkage over all subjects and iterations was 90.3\% with a common noise variance; 85.3\% with individual noise variance; 90.6\% with scaled noise variance; and 73.0\% with a global noise variance.  Using test-retest data to perform shrinkage, the median degree of shrinkage was 73.5\% with a common noise variance; 64.0\% with individual noise variance; 74.2\% with scaled noise variance; and 73.7\% with a global noise variance.

Figure \ref{fig:simulation1b} shows the MSE of the raw and shrinkage correlation estimates.  The average improvement in MSE due to shrinkage was fairly uniform across shrinkage methods under the default simulation parameters.  However, when a single session was used to perform shrinkage, there were more large outliers than when test-retest data was used, except when the global noise variance estimator was employed.  The median MSE over all subjects and iterations of the raw correlation estimates was 0.00498.  Using a single session to perform shrinkage, the median MSE of the shrinkage correlation estimates was 0.00130 (73.9\% lower) with a common noise variance; 0.00150 (69.9\% lower) with individual noise variance; 0.00131 (73.7\% lower) with scaled noise variance; and 0.00130 (73.9\% lower) with a global noise variance.  Using test-retest data to perform shrinkage, the median MSE of the shrinkage correlation estimates was 0.00119 (76.1\% lower) with a common noise variance; 0.00134 (73.1\% lower) with individual noise variance; 0.00118 (76.3\% lower) with scaled noise variance; and 0.00121 (75.7\% lower) with a global noise variance.

Figure \ref{fig:simulation1c} shows the Dice coefficient of similarity with the true parcellations of the parcellations generated from the raw and shrinkage correlation estimates.  The improvement in Dice coefficient due to shrinkage was fairly uniform across shrinkage methods but was maximized when test-retest data was used to perform shrinkage and when the global noise variance estimator was used with single session data.  The median Dice coefficient over all subjects and iterations of the raw parcellations was 0.750.  Using a single session to perform shrinkage, the median Dice coefficient of the shrinkage-based parcellations was 0.924 (23.2\% higher) with a common noise variance; 0.923 (23.1\% higher) with individual noise variance; 0.924 (23.2\% higher) with scaled noise variance; and 0.961 (28.1\% higher) with a global noise variance.  Using test-retest data to perform shrinkage, the median Dice coefficient of the shrinkage-based parcellations was 0.962 (28.3\% higher) with a common noise variance; 0.961 (28.1\% higher) with individual noise variance; 0.962 (28.3\% higher) with scaled noise variance; and 0.962 (28.3\% higher) with a global noise variance.

\begin{figure}
\begin{subfigure}[b]{0.5\textwidth}
\centering
\includegraphics[scale=0.32]{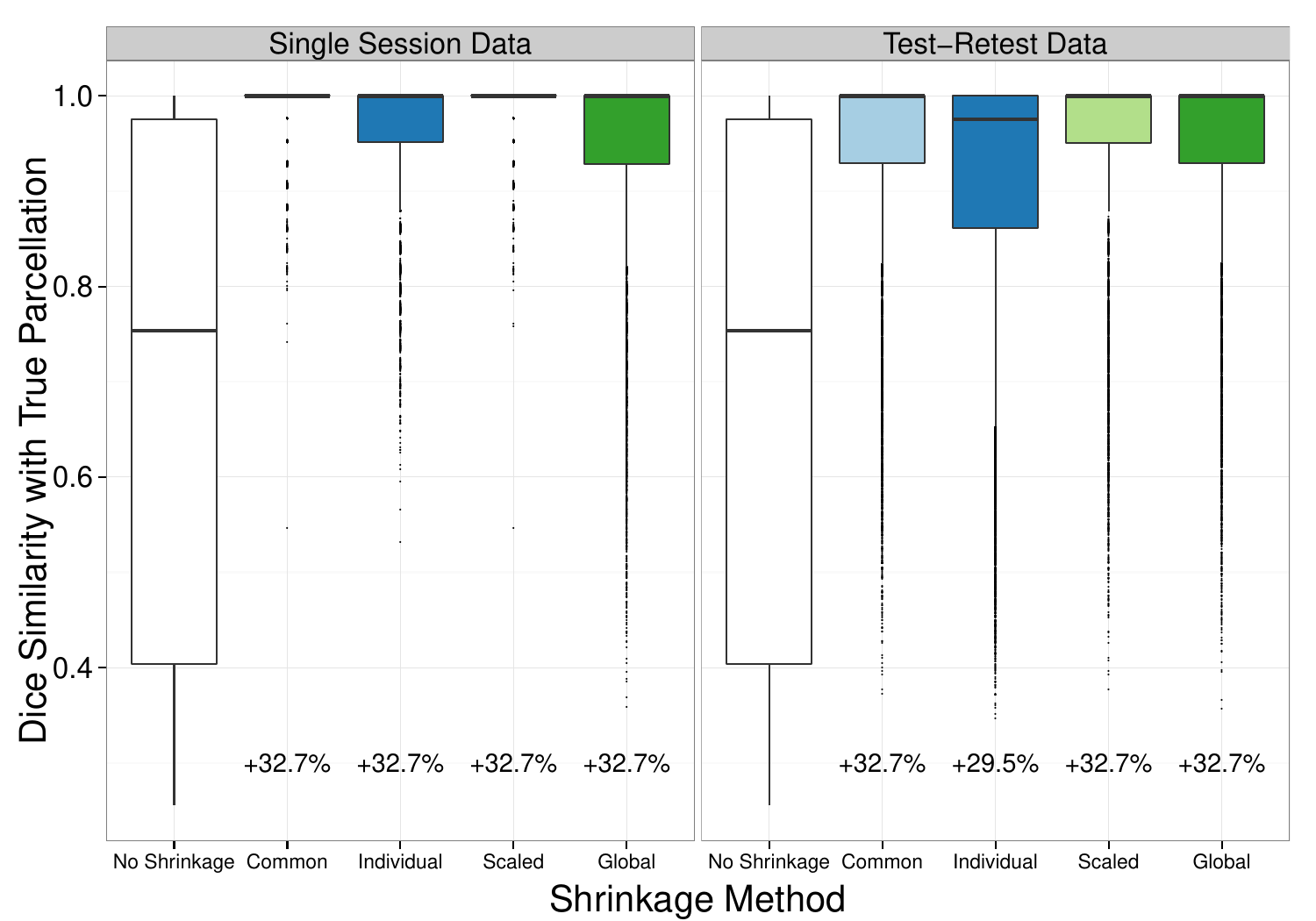}
\caption{Region where subjects have the same parcellation}
\label{fig:Dice_same}
\end{subfigure}
\begin{subfigure}[b]{0.5\textwidth}
\centering
\includegraphics[scale=0.32]{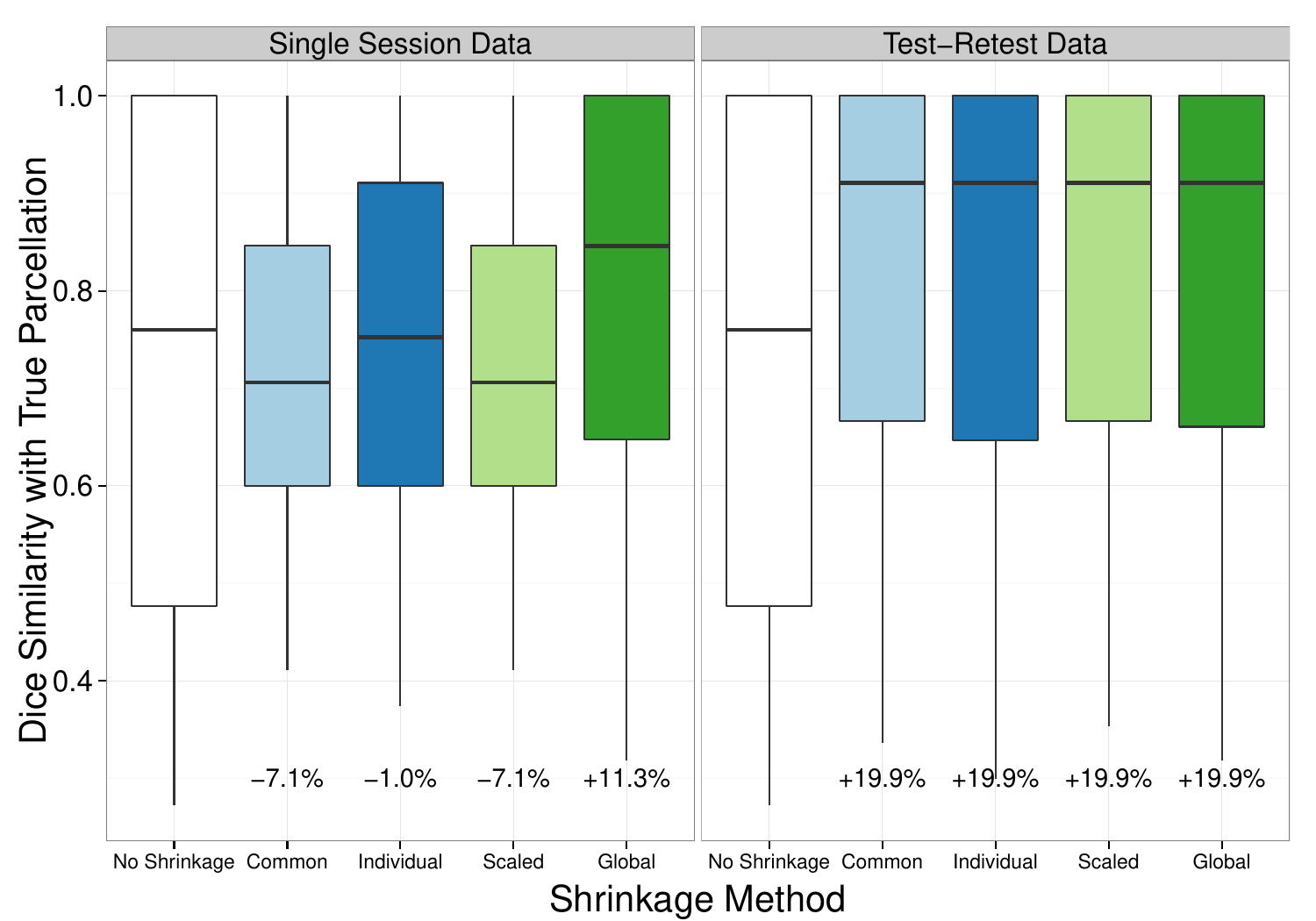}
\caption{Region where subjects' parcellations differ}
\label{fig:Dice_diff}
\end{subfigure}
\caption{Dice similarity (with each subject's true parcellation) within two different regions: the region of the image where all subjects share the same parcellation (a), and the region of the image where subjects' parcellations differ (b).  Results were computed under the default simulation parameters of $I=20$, $T=200$, $\rho=0.05$, $\sigma^2_X=0.02$, using either single session or test-retest data.}
\label{fig:Dice_same_diff}
\end{figure}

Figure \ref{fig:Dice_same_diff} illustrates the performance of the shrinkage-based parcellations in two different regions of the images displayed in Figure \ref{ClusterImg}.  The first is the region in which all subjects share the same parcellation, namely the top four rows and bottom four rows of the image.  In this region, shrinkage towards the group mean will clearly be beneficial, since the group mean is representative of the truth for each subject.  The second is the region in which subject-level differences in parcellations are allowed to occur, namely the middle two rows of the image.  In this region, it is less clear whether shrinkage will result in parcellations that are closer to the true subject-level parcellations.  In Figure \ref{fig:Dice_same}, we see that the improvement in the Dice coefficient within the first region was large for all shrinkage methods.  We also see that the methods that use a single session resulted in the greatest improvement, since these methods tend to over-estimate the noise variance and thus over-shrink.  In this region, since all subjects have the exact same parcellation, total shrinkage towards the group mean will be the most beneficial.  In Figure \ref{fig:Dice_diff}, as expected, we see that the improvement in the Dice coefficient within the second region was less dramatic.  In fact, for the shrinkage methods that tend to over-shrink, there was a reduction in the median Dice coefficient compared with the raw parcellations.  However, when test-retest data was used to perform shrinkage or the global noise variance estimator was used with single session data, there was an improvement in the median Dice coefficient.  When test-retest data was used to perform shrinkage, the median Dice coefficient increased by 19.9\% due to shrinkage for all noise variance estimators; when a single session was used to perform shrinkage and the global noise variance estimator was used, the median Dice coefficient increased by 11.3\% due to shrinkage.

\textbf{Analysis S2: Sensitivity to simulation parameters}

\begin{figure}
\begin{subfigure}[b]{1\textwidth}
\centering
\includegraphics[scale=0.3, trim=0 15mm 7.5mm 0, clip]{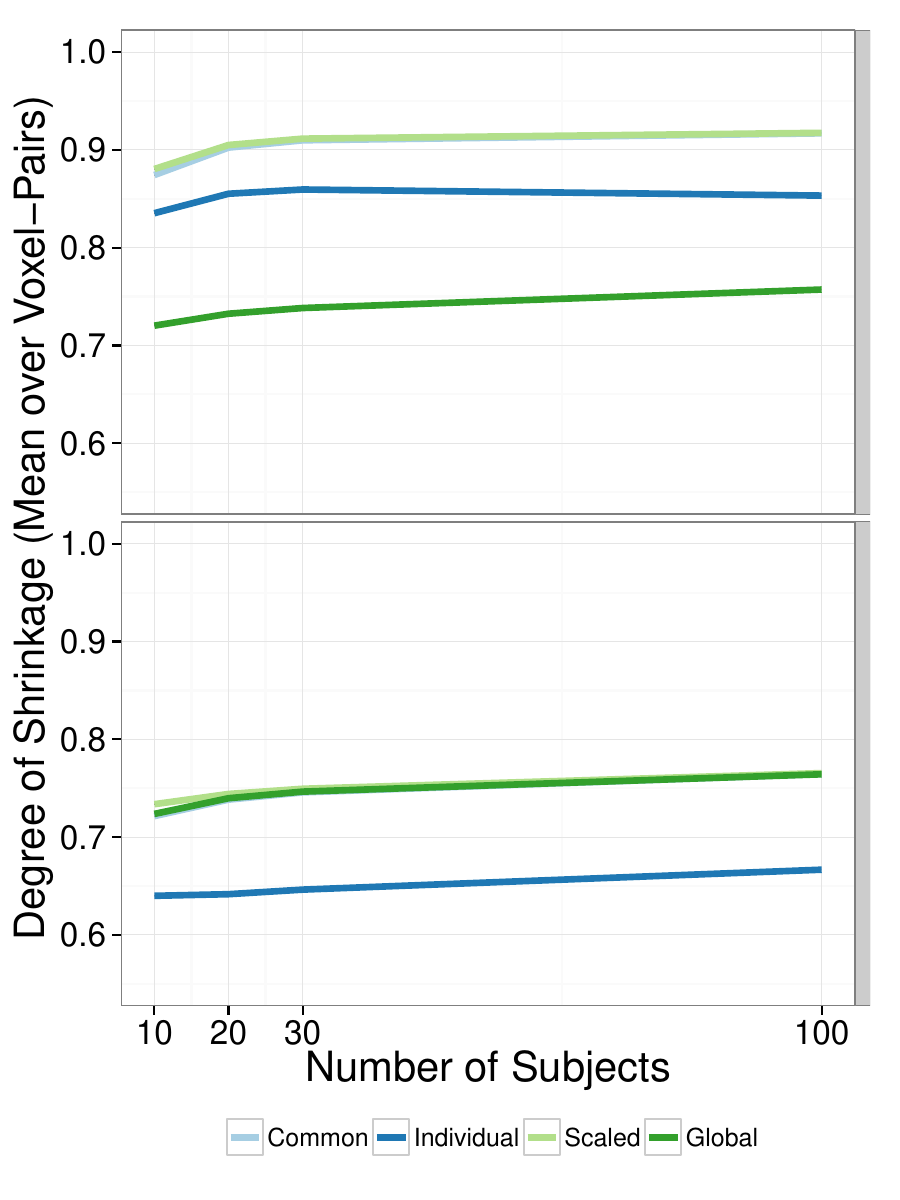}
\includegraphics[scale=0.3, trim=12mm 15mm 7.5mm 0, clip]{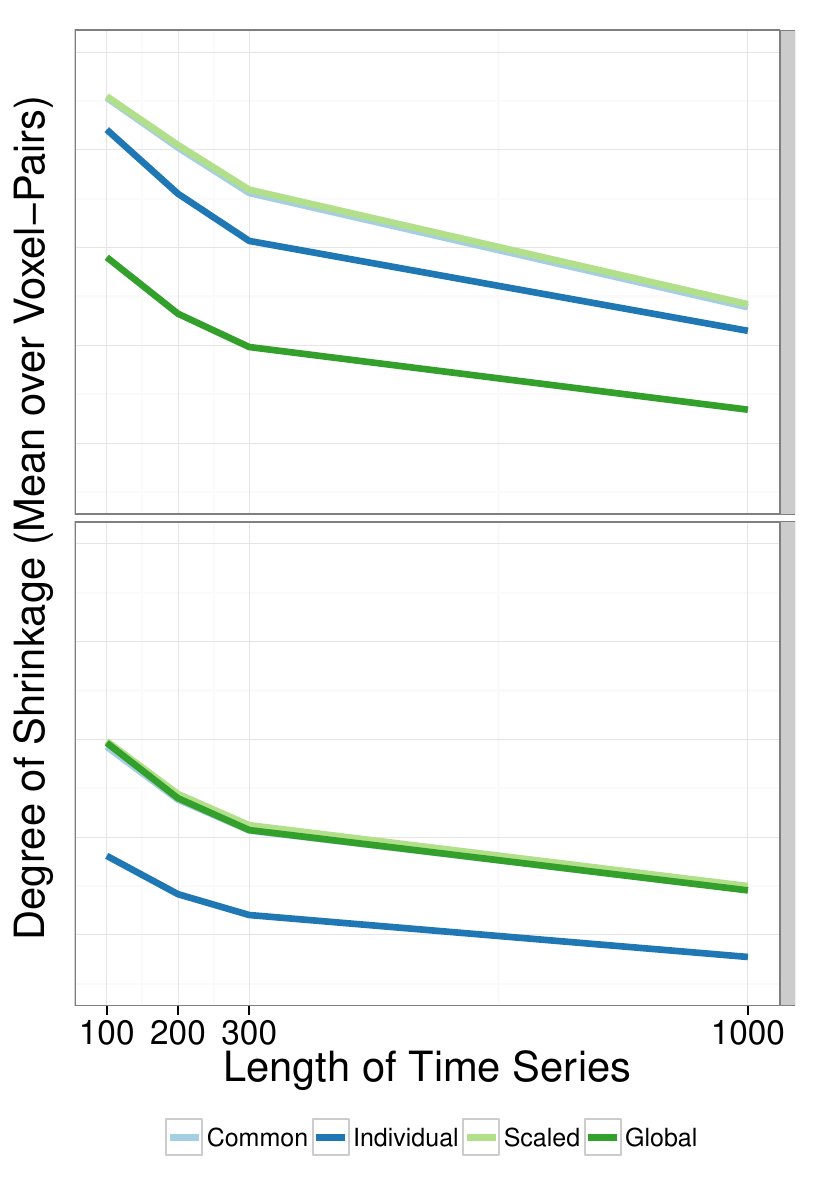}
\includegraphics[scale=0.3, trim=12mm 15mm 7.5mm 0, clip]{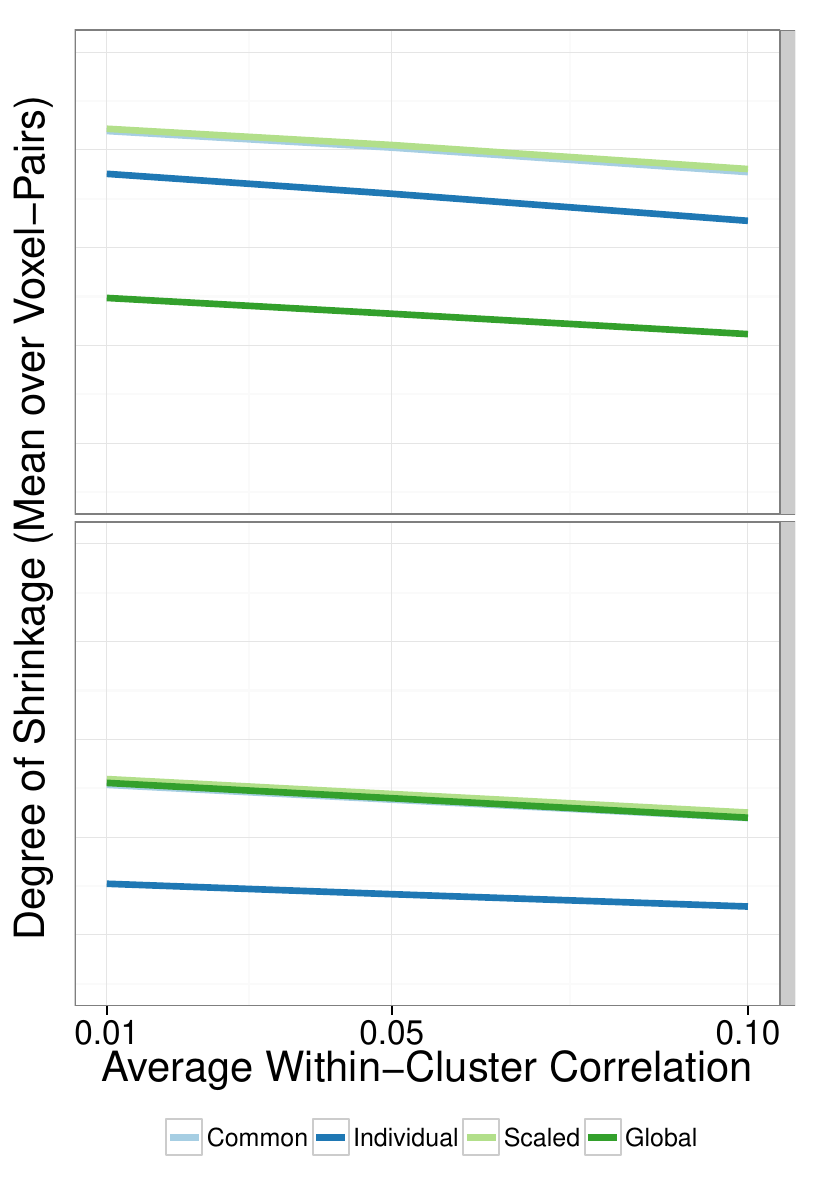}
\includegraphics[scale=0.3, trim=12mm 15mm 4mm 0, clip]{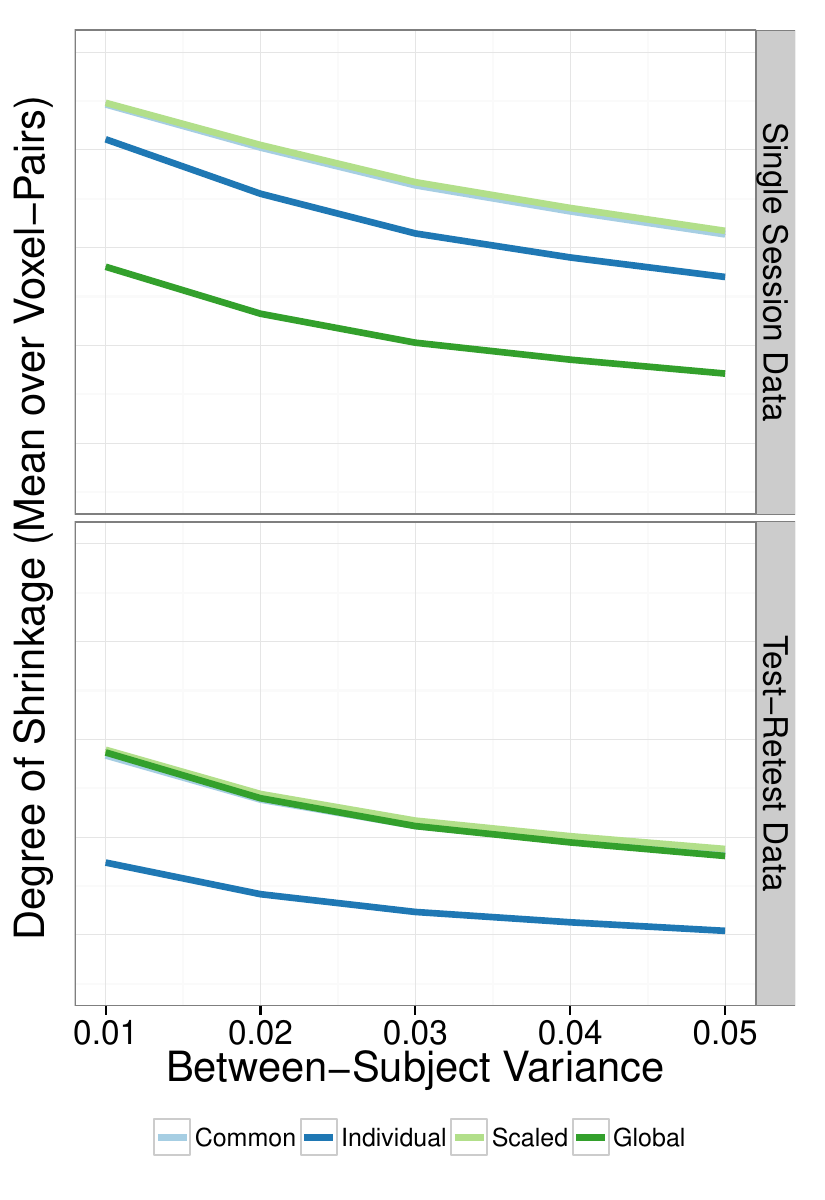}\\
\includegraphics[scale=0.5]{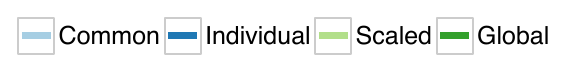}\\
\caption{Effect of simulation parameters on degree of shrinkage\\[10pt]}
\label{fig:simulation_lambda}
\end{subfigure}
\begin{subfigure}[b]{1\textwidth}
\centering
\includegraphics[scale=0.3, trim=0 15mm 7.5mm 0, clip]{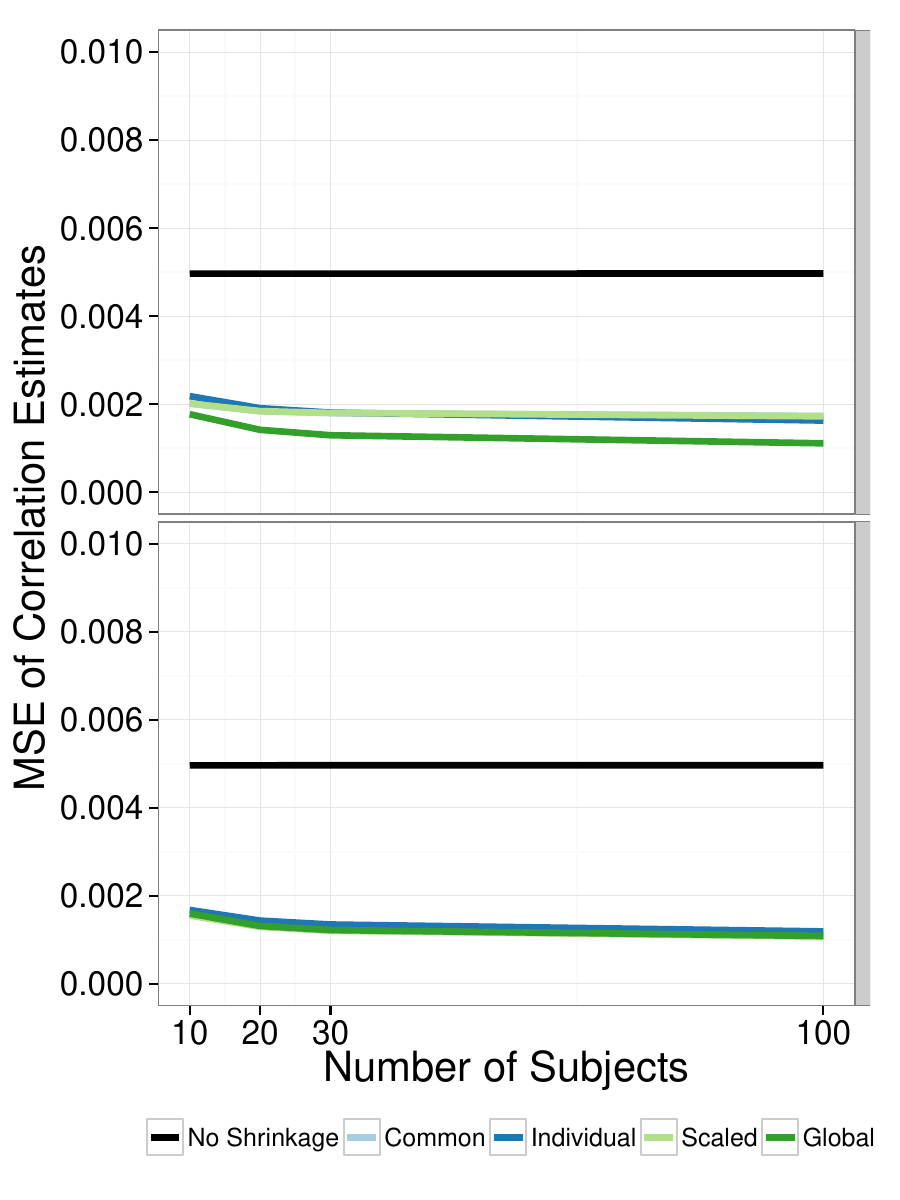}
\includegraphics[scale=0.3, trim=12mm 15mm 7.5mm 0, clip]{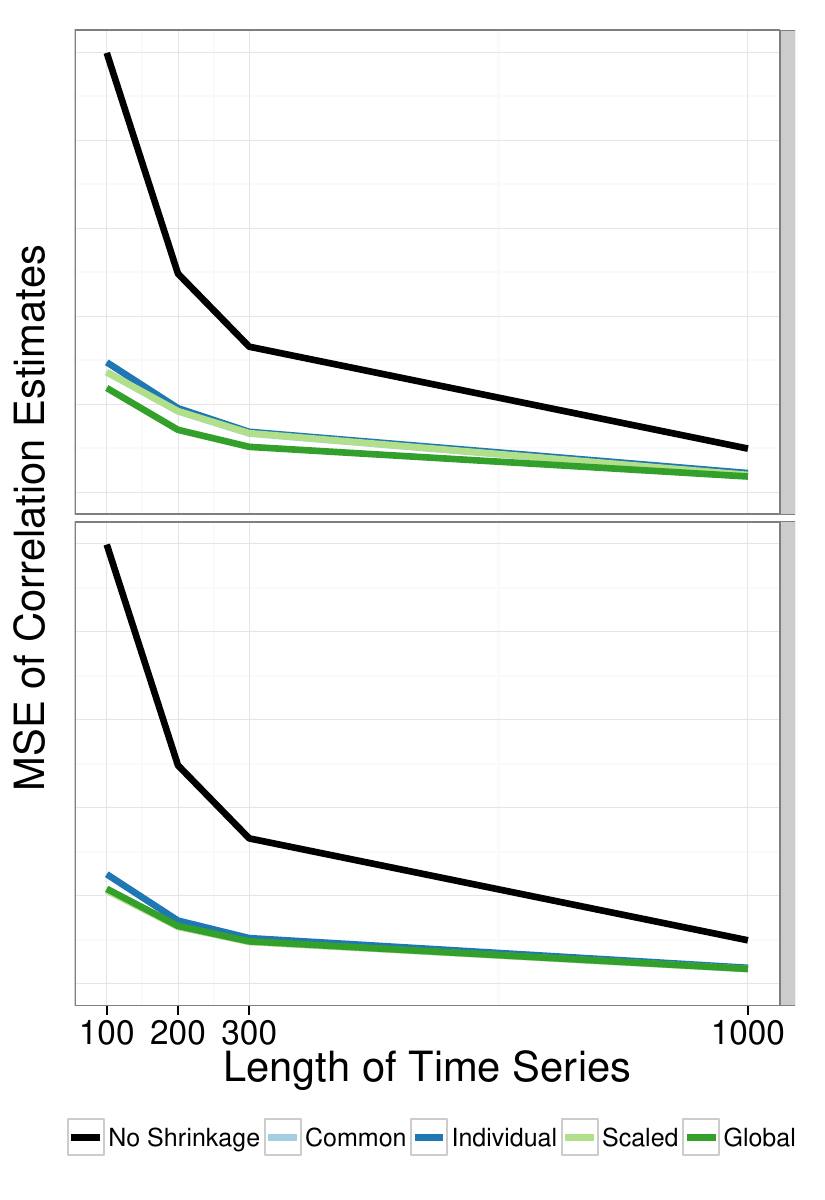}
\includegraphics[scale=0.3, trim=12mm 15mm 7.5mm 0, clip]{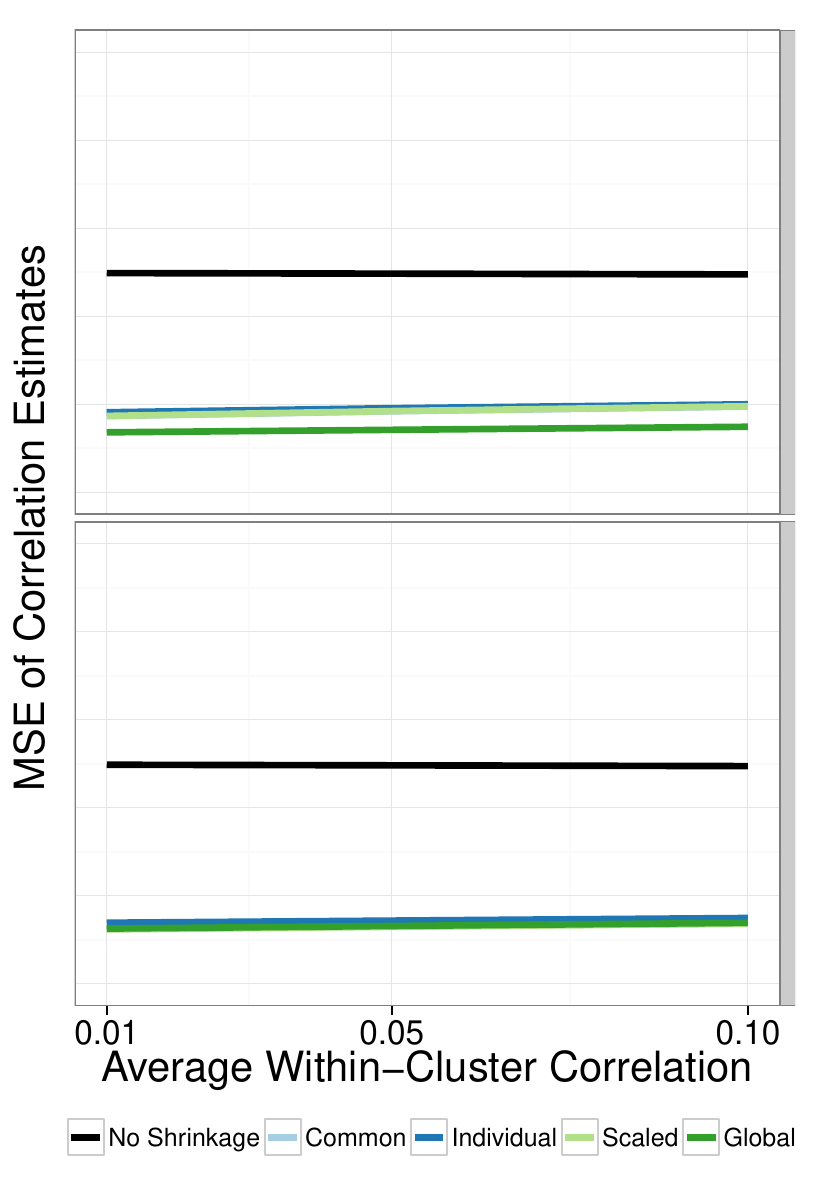}
\includegraphics[scale=0.3, trim=12mm 15mm 4mm 0, clip]{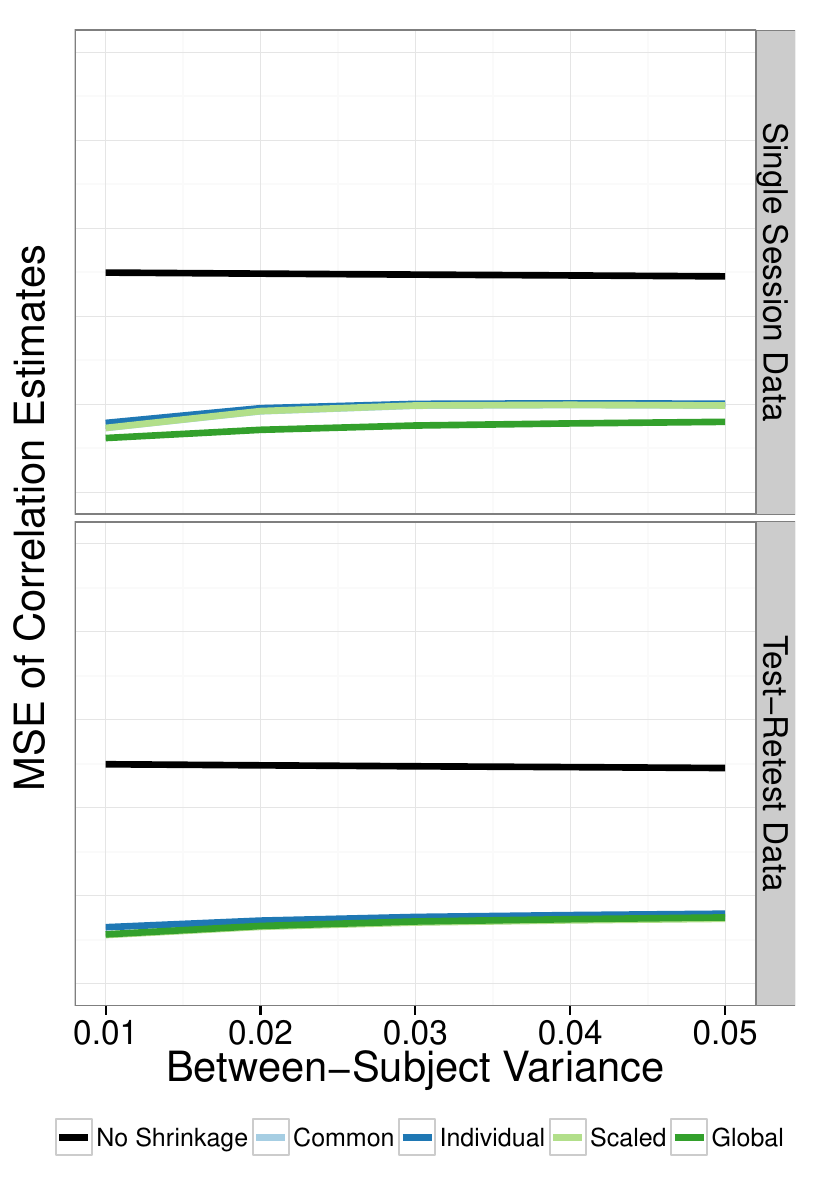}\\
\includegraphics[scale=0.5]{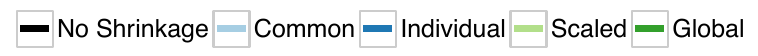}\\
\caption{Effect of simulation parameters on test-retest error of correlation estimates\\[10pt]}
\label{fig:simulation_MSE}
\end{subfigure}
\begin{subfigure}[b]{1\textwidth}
\centering
\includegraphics[scale=0.3, trim=0 15mm 7.5mm 0, clip]{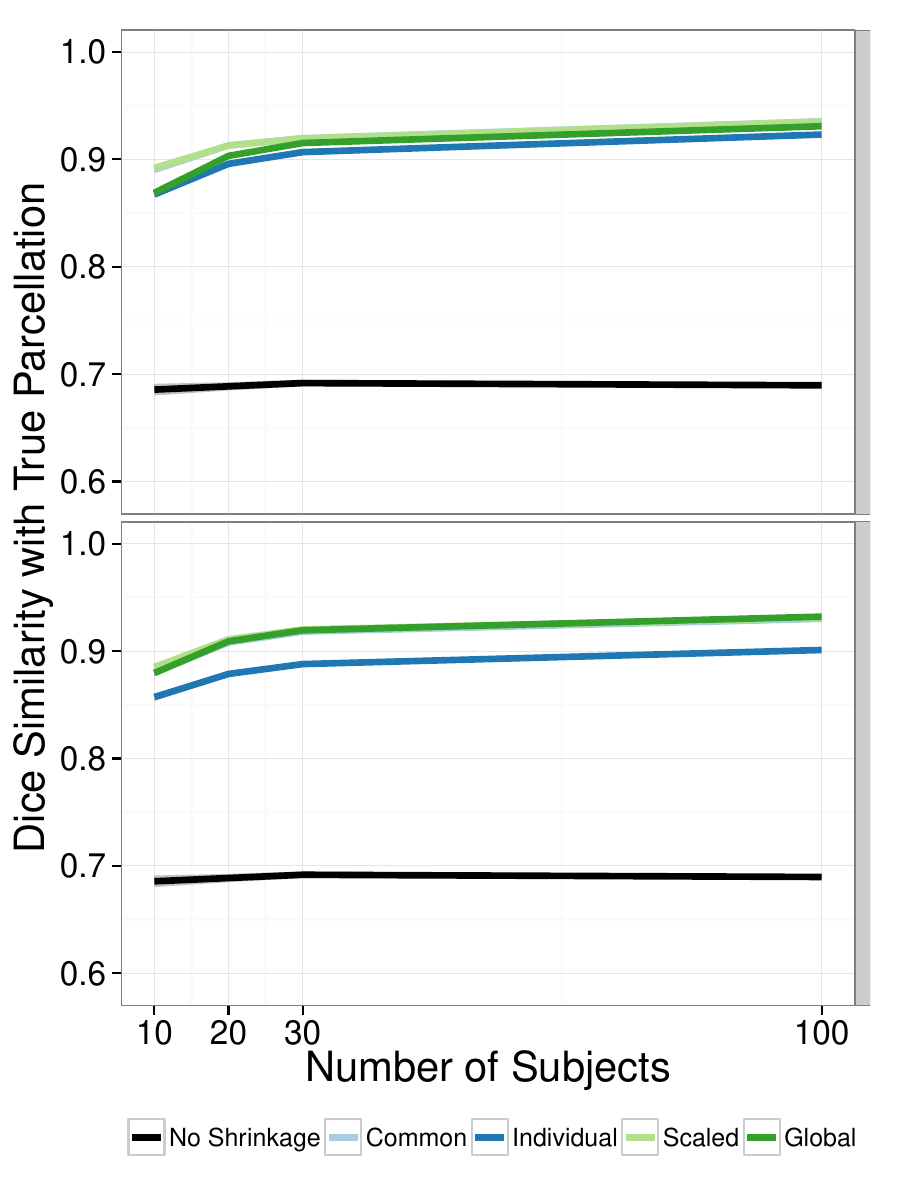}
\includegraphics[scale=0.3, trim=12mm 15mm 7.5mm 0, clip]{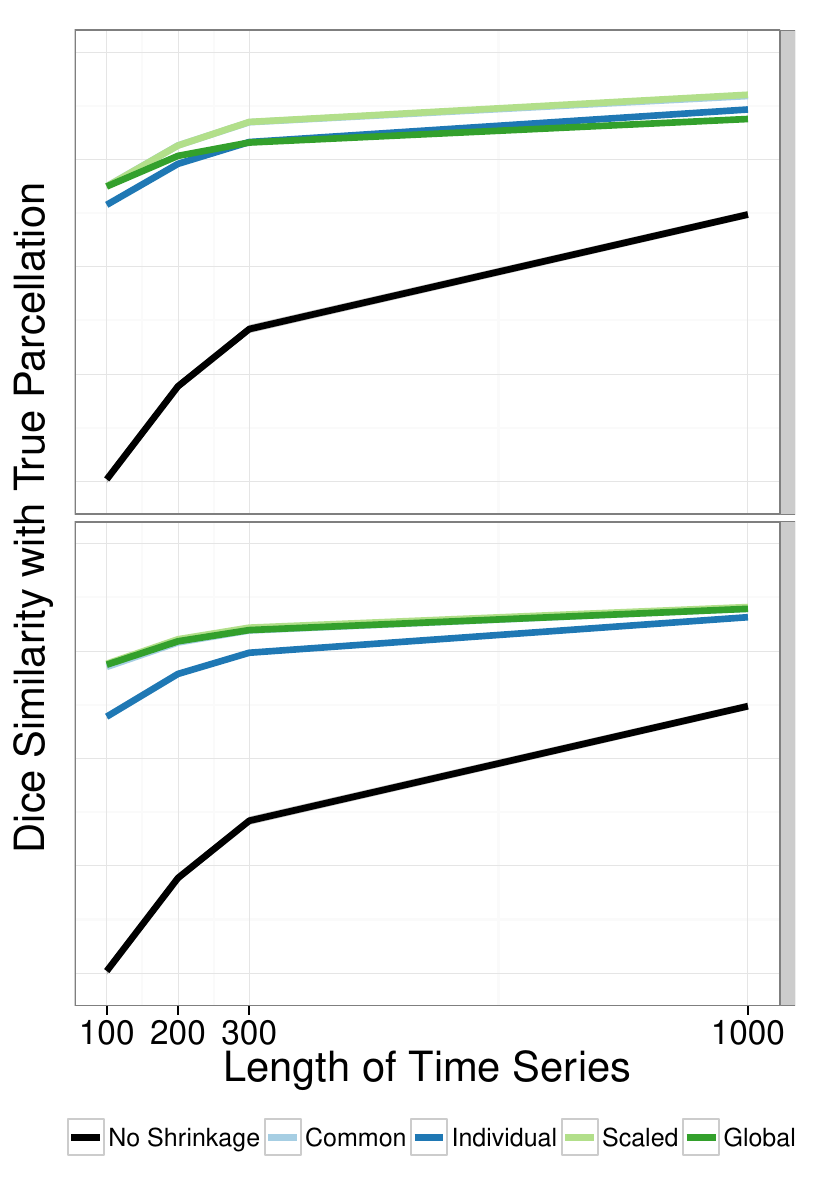}
\includegraphics[scale=0.3, trim=12mm 15mm 7.5mm 0, clip]{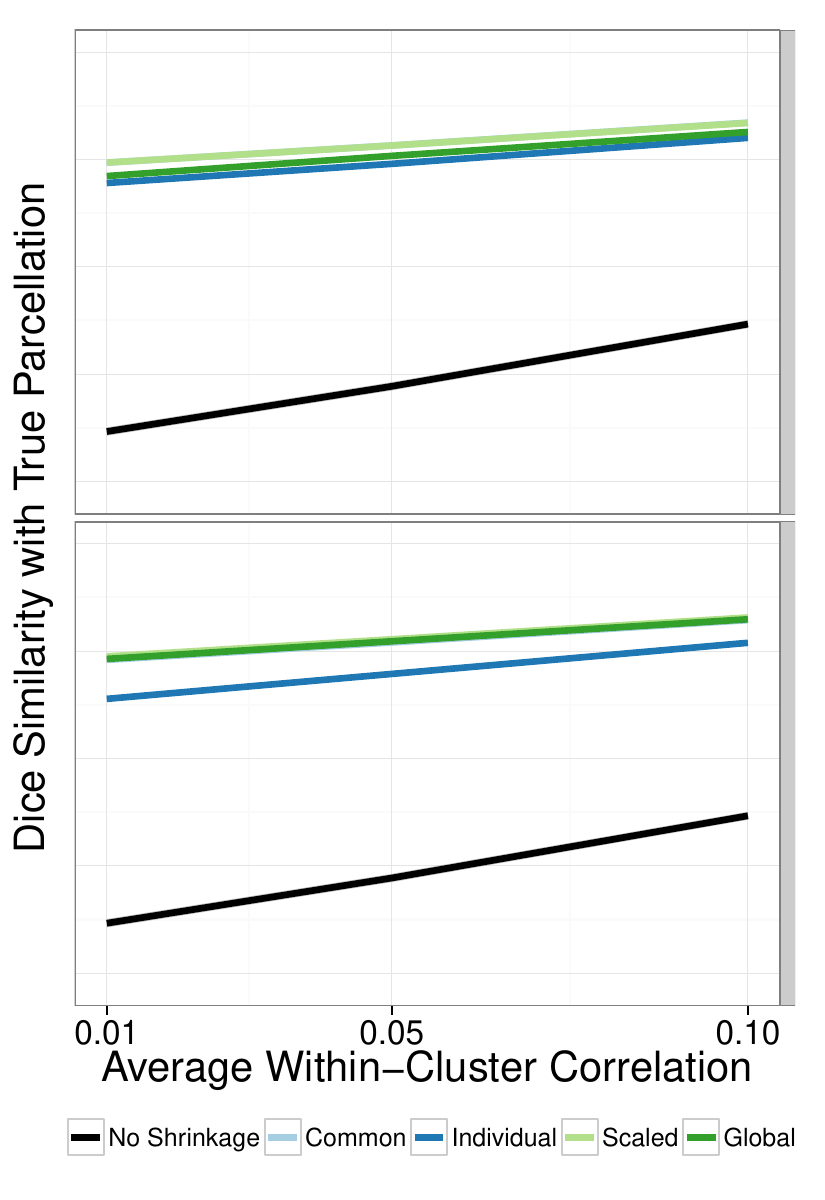}
\includegraphics[scale=0.3, trim=12mm 15mm 4mm 0, clip]{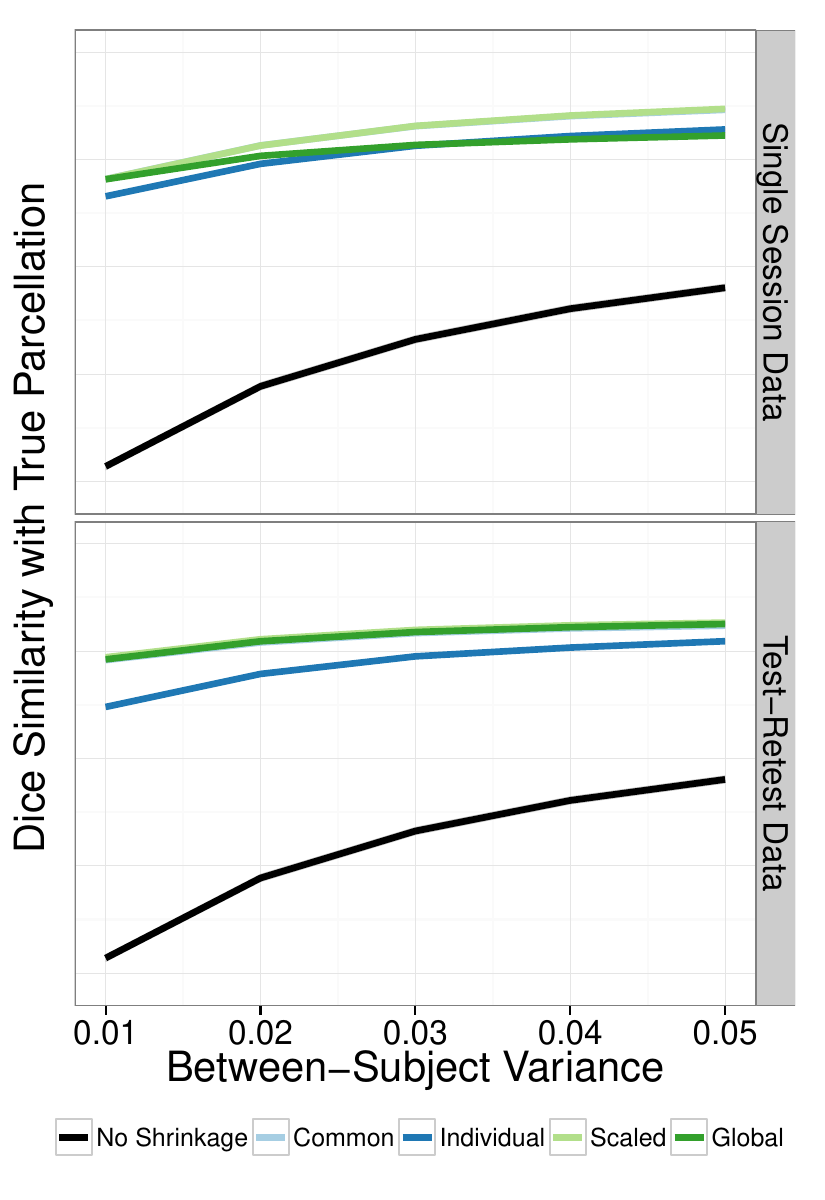}\\
\includegraphics[scale=0.5]{figures/scale_Dice.pdf}\\
\caption{Effect of simulation parameters on test-retest reliability of parcellations\\[10pt]}
\label{fig:simulation_Dice}
\end{subfigure}
\caption{Effect of each simulation parameter on the degree of shrinkage, MSE of the correlation estimates, and Dice similarity of the subject-level parcellations.  Each point shows the median value over all subjects and iterations.  Error bands show $\pm 2$ standard errors around the median (may not be visible due to narrow width).}
\label{fig:simulation2}
\end{figure}

Figure \ref{fig:simulation2} shows how the degree of shrinkage (Figure \ref{fig:simulation_lambda}), the MSE of the correlation estimates (Figure \ref{fig:simulation_MSE}), and the Dice similarity of the parcellations (Figure \ref{fig:simulation_Dice}) change with varying simulation parameters.  The first column shows the effect of varying the number of subjects; the second column shows the effect of varying the time series length; the third column shows the effect of varying the strength of inter-voxel correlations; and the fourth column shows the effect of varying the degree of similarity across subjects.  Each color represents a different shrinkage method, and results are shown in the case of both single session data (top panel of each plot) and test-retest data (bottom panel of each plot).  Each point represents the mean over all subjects and iterations, and 95\% confidence intervals are shown as grey bands, which may not be visible due to their narrow width.

Figure \ref{fig:simulation_lambda} shows that the degree of shrinkage tends to decrease as the time series length $T$ increases, as the within-cluster correlation $\rho$ increases, and as the between-subject variance $\sigma^2_X$ increases.  This is expected, since the shrinkage parameter is defined as the ratio of within-subject variance to total (within-subject plus between-subject) variance, and higher values of $T$ and $\rho$ reduce the within-subject variance.  There is a weak increase in the degree of shrinkage as the number of subjects $I$ increases.  This reflects bias in the estimation of $\lambda_i(v,v')$, a non-linear function of variance components, a bias that diminishes as the sample size increases.

Figure \ref{fig:simulation_MSE} shows that the MSE of the raw estimates is primarily related to the time series length $T$, since as $T$ increases, sampling variability decreases.  As $T$ increases, the MSE of all estimators decreases, and the MSE of the raw estimator approaches, but does not achieve, the MSE of the shrinkage estimators.  As the number of subjects $I$ increases, there is also a reduction in the MSE of the shrinkage estimators, which is due to the increase in the degree of shrinkage associated with larger sample size.

Figure \ref{fig:simulation_Dice} shows that the Dice coefficient of the raw parcellations increases as the time series length $T$ increases, as the within-cluster correlation $\rho$ increases, and as the between-subject variance $\sigma^2_X$ increases.  Similar to the results for MSE of the correlation estimates, as $T$ increases, the Dice coefficient of the raw parcellations approaches that of the shrinkage-based parcellations; unlike the results for MSE, the shrinkage-based parcellations still dramatically outperform the raw parcellations even at $T=1000$.  The Dice coefficient of the shrinkage-based parcellations increases along with the Dice coefficient of the raw parcellations as $T$, $\rho$, or $\sigma^2_X$ is increased.  As the number of subjects $I$ increases, the Dice coefficient of the shrinkage estimators increases, which is again due to the increase in the degree of shrinkage associated with larger sample size.

\subsection{Real fMRI Dataset Results}

\begin{figure}
\begin{subfigure}[b]{0.5\textwidth}
\centering
\includegraphics[scale=0.32]{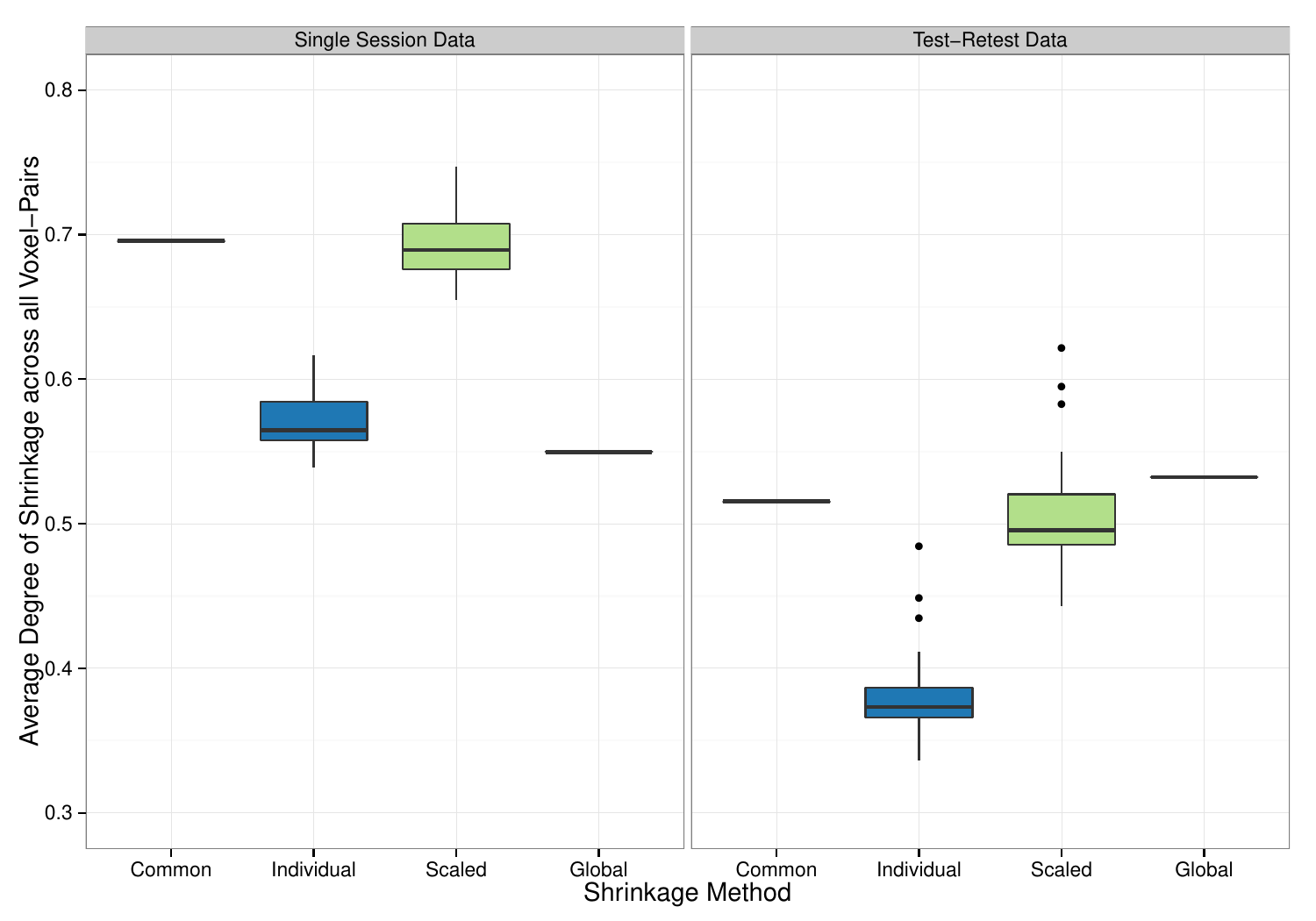}
\caption{Shrinkage on Fisher-transformed correlations}
\label{fig:lambda_kirby_Z}
\end{subfigure}
\begin{subfigure}[b]{0.5\textwidth}
\centering
\includegraphics[scale=0.32]{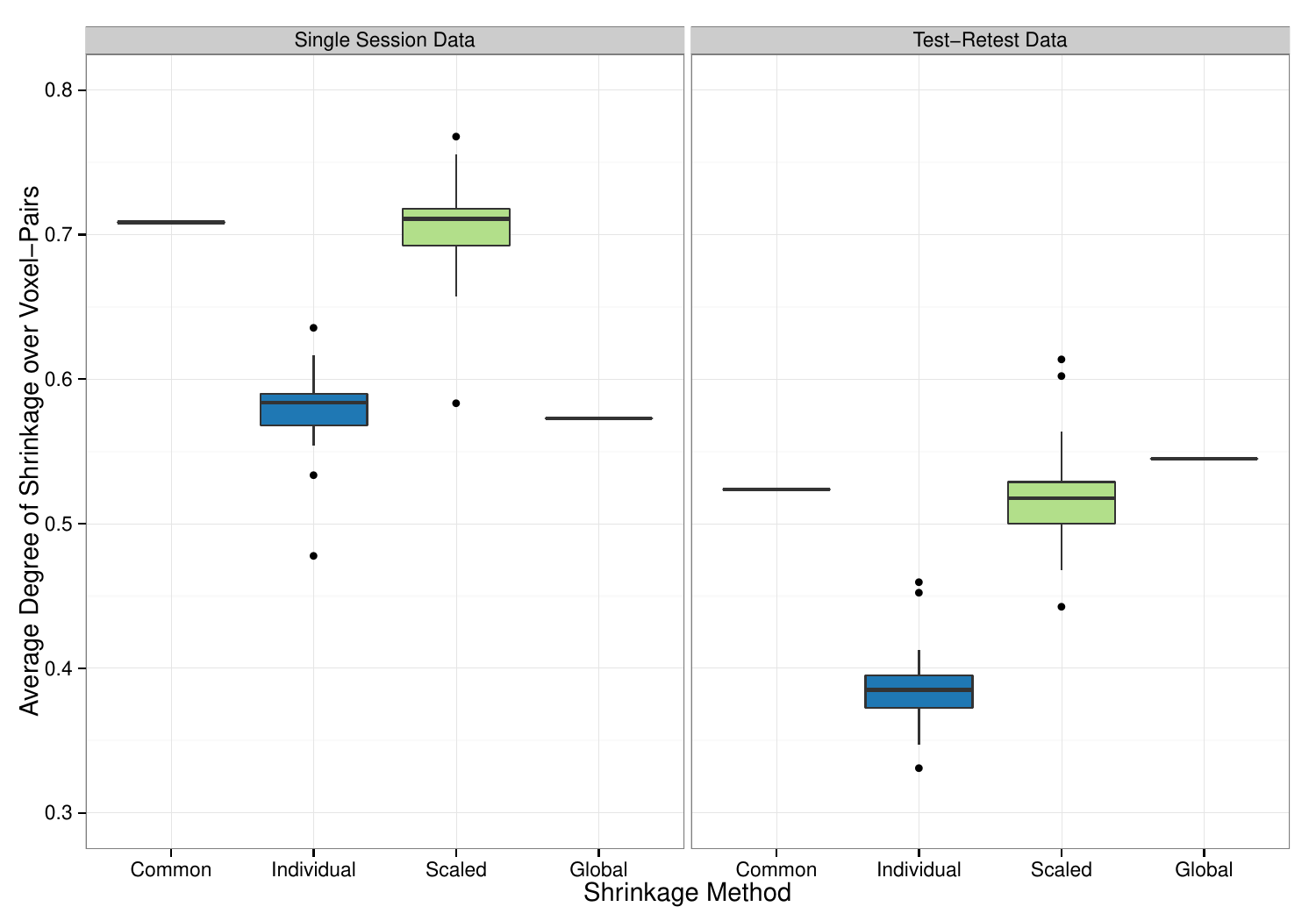}
\caption{Shrinkage on correlations}
\label{fig:lambda_kirby}
\end{subfigure}
\caption{Degree of shrinkage (percent weighting of the group mean, averaged over all voxel-pairs) by noise variance estimation method and type of dataset (single session or test-retest) used to perform shrinkage.  For the ``common'' and ``global'' noise variance methods, all subjects have the same shrinkage parameter at each voxel-pair, so the boxplot shows a single value.\\[12pt]}
\label{fig:lambda}
%\end{figure}

%\begin{figure}
\begin{subfigure}[b]{0.5\textwidth}
\centering
\includegraphics[scale=0.32]{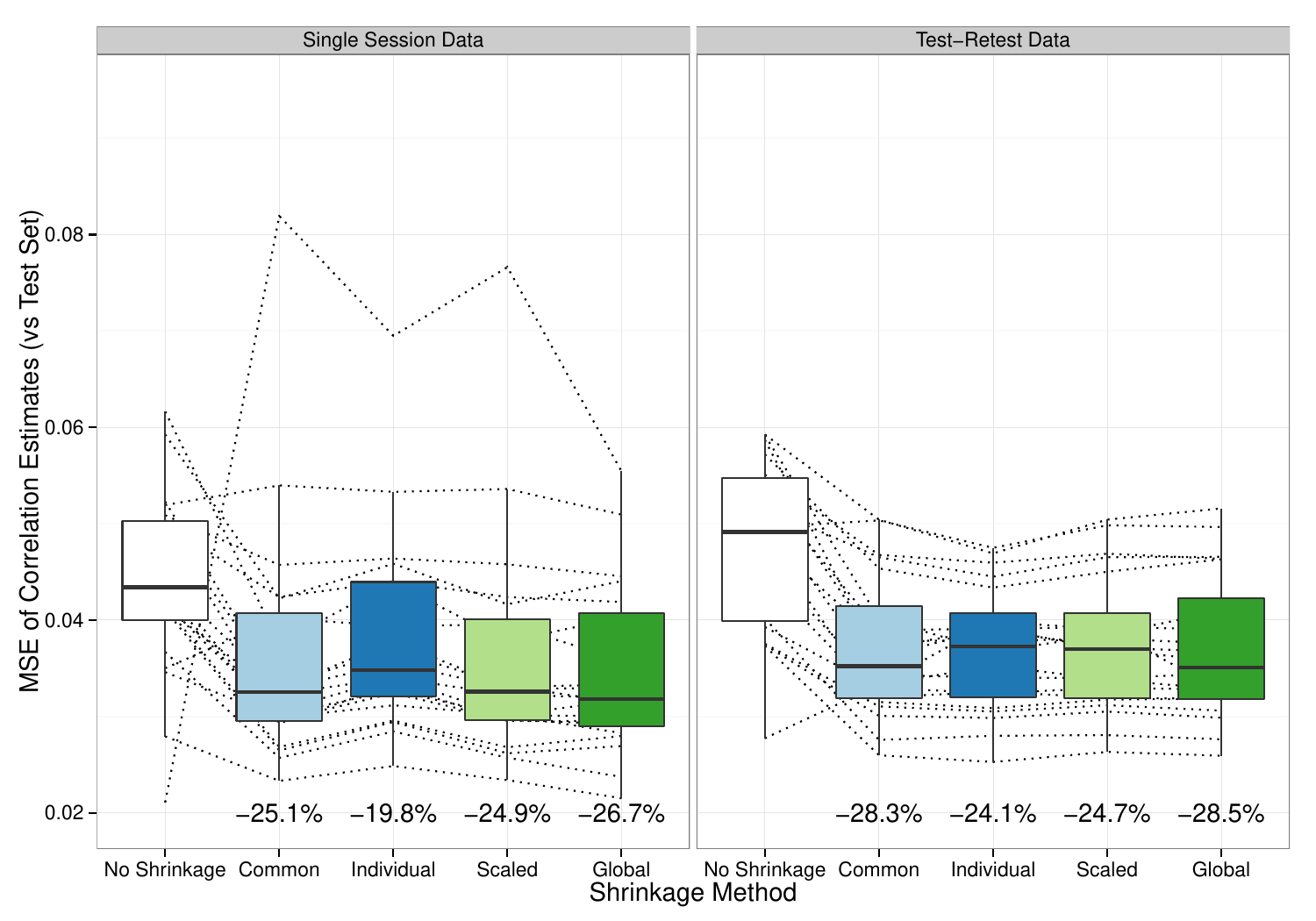}
\caption{Shrinkage on Fisher-transformed correlations}
\label{fig:MSE_kirby_Z}
\end{subfigure}
\begin{subfigure}[b]{0.5\textwidth}
\centering
\includegraphics[scale=0.32]{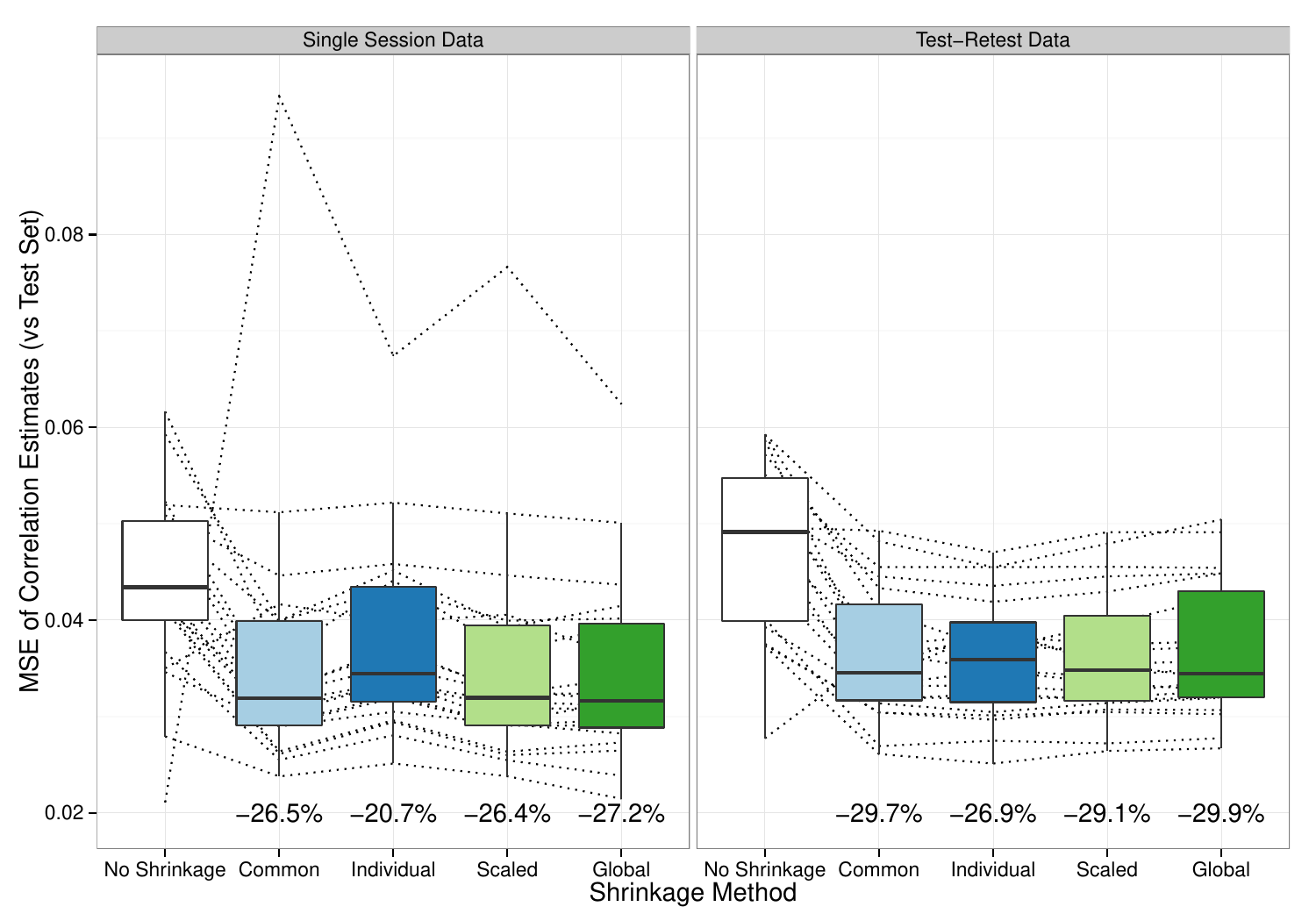}
\caption{Shrinkage on correlations}
\label{fig:MSE_kirby}
\end{subfigure}
\caption{MSE of raw and shrinkage correlation estimates, by noise variance estimation method and the type of dataset (single session or test-retest) used to perform shrinkage.  Each dotted line shows the MSE for a single subject's raw and shrinkage estimates, and the boxplots show the distributions over all subjects.  The percent decrease in the median MSE of each shrinkage estimate (compared to the median MSE of the raw estimate) is reported below each boxplot.}
\label{fig:MSE}
\end{figure}

\begin{figure}
\centering
\includegraphics[scale=0.52]{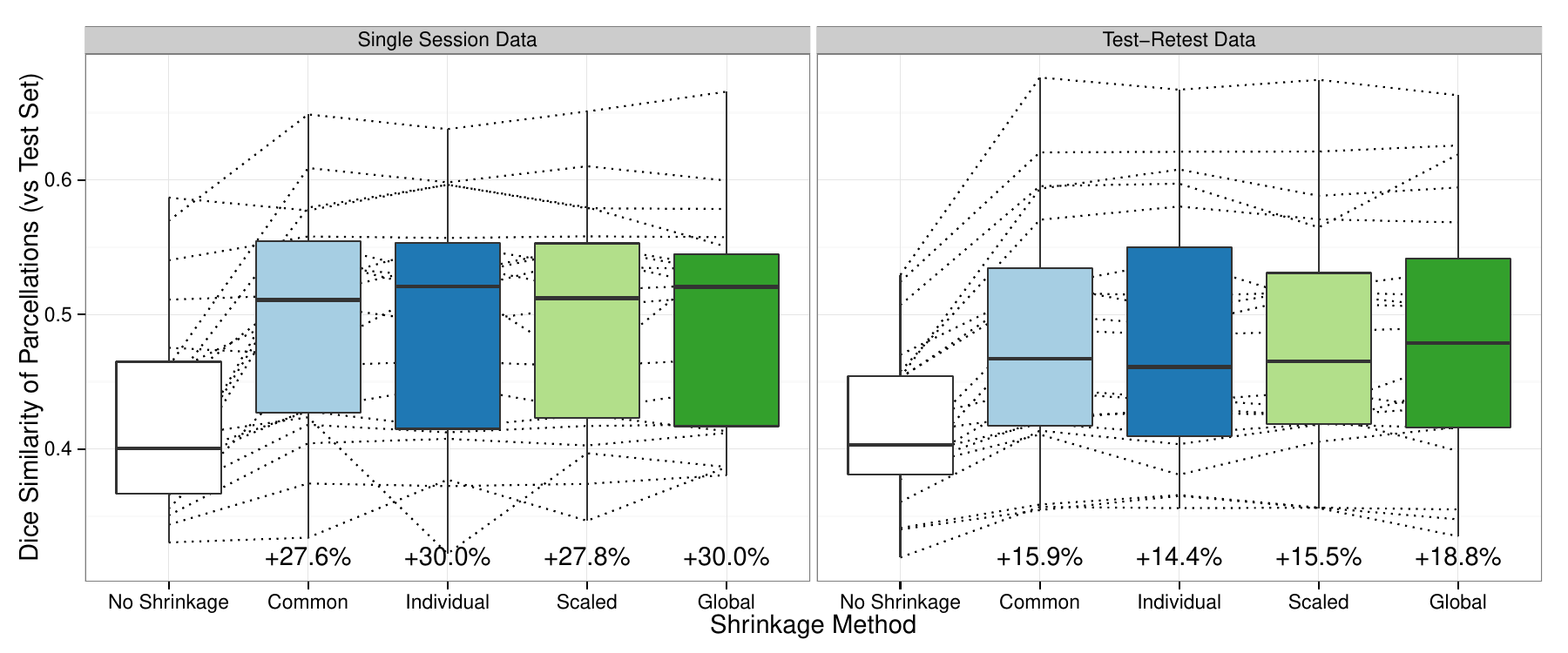}
\caption{Dice coefficients of similarity (with test set parcellations) of parcellations produced using raw and shrinkage correlation estimates.  Results are shown by noise variance estimation method and the type of dataset (single session or test-retest) used to perform shrinkage.  Each dotted line shows the Dice coefficients for a single subject, and the boxplots show the distributions over subjects.  The percent increase in the median Dice coefficient of each shrinkage parcellation (compared to the median Dice coefficient of the raw parcellation) is reported below each boxplot.}
\label{fig:Dice}
\end{figure}

Figure \ref{fig:lambda} shows the degree of shrinkage towards the group mean performed on the real rsfMRI dataset by noise variance estimation method and the type of dataset (single session or test-retest) used to perform shrinkage.  As in the simulation results, the degree of shrinkage for subject $i$ was computed as the mean value of the shrinkage parameter $\lambda_i(v,v')$ over all voxel-pairs $(v,v')$.  Each boxplot shows the distribution of these values over subjects.  For those methods that computed only a group-level shrinkage parameter (the common and global noise variance methods), the shrinkage parameter is the same over all subjects and the boxplot shows only a single value.

Below, results are reported for the case of shrinkage on Fisher-transformed correlations, followed in brackets by results for the case of shrinkage directly on correlations.  

When a single session was used to compute the noise variance, the median degree of shrinkage over all subjects was 69.6\% [70.8\%] with a common noise variance; 56.5\% [58.4\%] with individual noise variance; 68.9\% [71.1\%] with scaled noise variance; and 55.0\% [57.3\%] with a global noise variance.  When test-retest data was used to compute the noise variance, the median degree of shrinkage over all subjects was 51.5\% [52.4\%] with a common noise variance; 37.3\% [38.5\%] with individual noise variance; 49.5\% [51.8\%] with scaled noise variance; and 53.2\% [54.5\%] with a global noise variance.  

As expected, the degree of shrinkage was generally higher when a single session was used due to upward bias in the noise variance estimation.  As the global noise variance estimator was designed to avoid this problem it does not suffer from inflated degree of shrinkage.  By contrast, when the individual noise variance estimator was used, the degree of shrinkage was significantly lower compared with the common or scaled noise variance estimator.  Since each individual noise variance estimate $\hat\sigma^2_i(v,v')$ is based on only two observations (rather than $S=20$), the distribution around the truth $\sigma^2_i(v,v')$ is a highly skewed Chi-squared, which introduces bias into the shrinkage parameter estimator, since it is a non-linear function of the variance estimators. 

\textbf{Analysis R1: Performance of shrinkage estimates}

Figure \ref{fig:MSE} shows the MSE of each raw and shrinkage correlation estimate by the noise variance estimation method employed and the type of dataset (single session or test-retest) used to perform shrinkage.  Each dotted line represents a single subject, and the boxplots show the distribution of values over all subjects.  Below each boxplot we report the percent decrease in the median MSE of the shrinkage estimates compared to the median MSE of the raw estimates.  Figure \ref{fig:MSE_kirby_Z} shows the results from applying shrinkage on the Fisher-transformed correlations, and Figure \ref{fig:MSE_kirby} shows the results from applying shrinkage directly to the untransformed correlations.  

Results are again reported for the case of shrinkage on Fisher-transformed correlations, followed in brackets by results for the case of shrinkage directly on correlations. 

All shrinkage methods resulted in a decrease in the median MSE compared with the raw estimates.  Recall that the data used to compute the raw estimates and parcellations was different in the single session case and the test-retest case (see Figure \ref{Data_setup}).  Specifically, the full 7-minute scan from the first session was used in the single session case, and only the first 4 minutes and 40 seconds of that scan was used in the test-retest case.  Therefore, the raw coefficients and parcellations, and their respective reliability measures, differ across the two cases.  When a single session was used to compute the noise variance, the raw correlation estimates had a median MSE of 0.0434.  The shrinkage estimates had a median MSE of 0.0325 (25.1\% lower) [0.0319 (26.5\% lower)] with a common noise variance; 0.0348 (19.8\% lower) [0.0344 (20.7\% lower)] with individual noise variance; 0.0326 (24.9\% lower) [0.0319 (26.4\% lower)] with scaled noise variance; and 0.0318 (26.7\% lower) [0.0316 (27.2\% lower)] with a global noise variance. When test-retest data was used to compute the noise variance, the raw correlation estimates had a median MSE of 0.0491.  The shrinkage estimates had a median MSE of 0.0352 (28.3\% lower) [0.0345 (29.7\% lower)] with a common noise variance; 0.0373 (24.1\% lower) [0.0359 (26.9\% lower)] with individual noise variance; 0.0370 (24.7\% lower) [0.0348 (29.1\% lower)] with scaled noise variance; and 0.0351 (28.5\% lower) [0.0344 (29.9\% lower)] with a global noise variance.

For all methods, applying shrinkage directly to the correlations resulted in greater reduction in MSE (compared with applying shrinkage to the Fisher-transformed correlations).  Whether single session data or test-retest data was used, shrinkage using the global noise variance estimator resulted in the greatest reduction in median MSE. At the subject level, when test-retest data was used to perform shrinkage, shrinkage on untransformed correlations resulted in reduced MSE for 19 out of 20 subjects across all noise variance estimation methods; when single session data was used, shrinkage on untransformed correlations resulted in reduced MSE for 17 out of 20 subjects for the individual noise variance estimation method and 18 out of 20 subjects for all other noise variance estimation methods.  

\begin{figure}
\begin{subfigure}[b]{1\textwidth}
\centering
\includegraphics[scale=0.35, trim=9cm 7cm 9cm 4cm, clip]{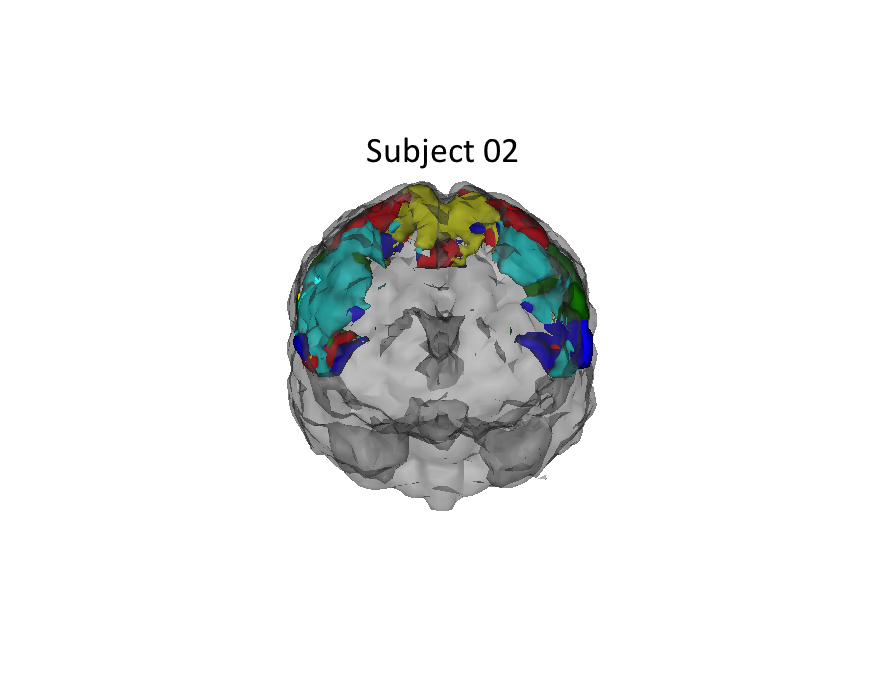}
\includegraphics[scale=0.35, trim=9cm 7cm 9cm 4cm, clip]{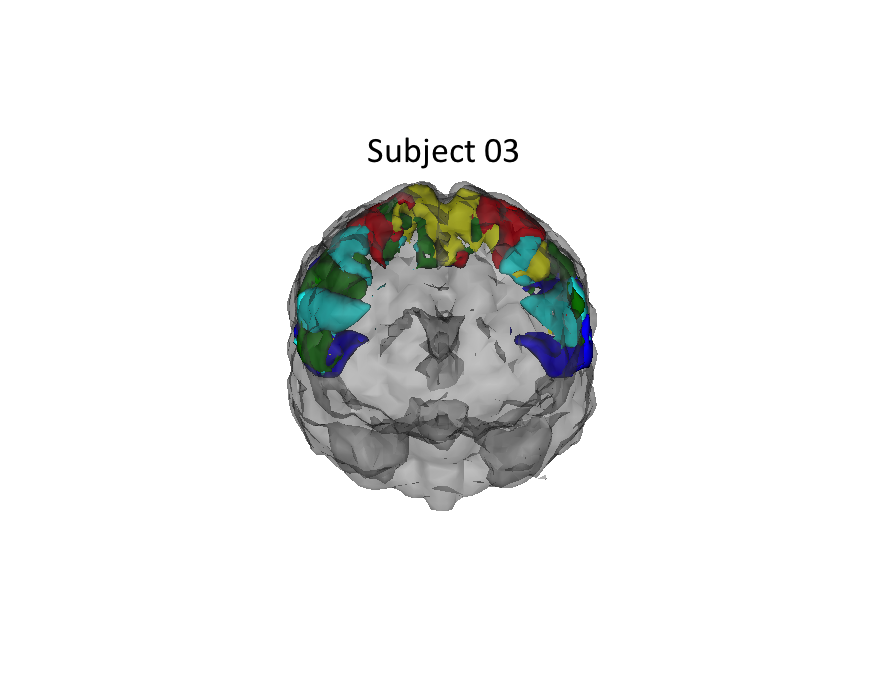}
\includegraphics[scale=0.35, trim=9cm 7cm 9cm 4cm, clip]{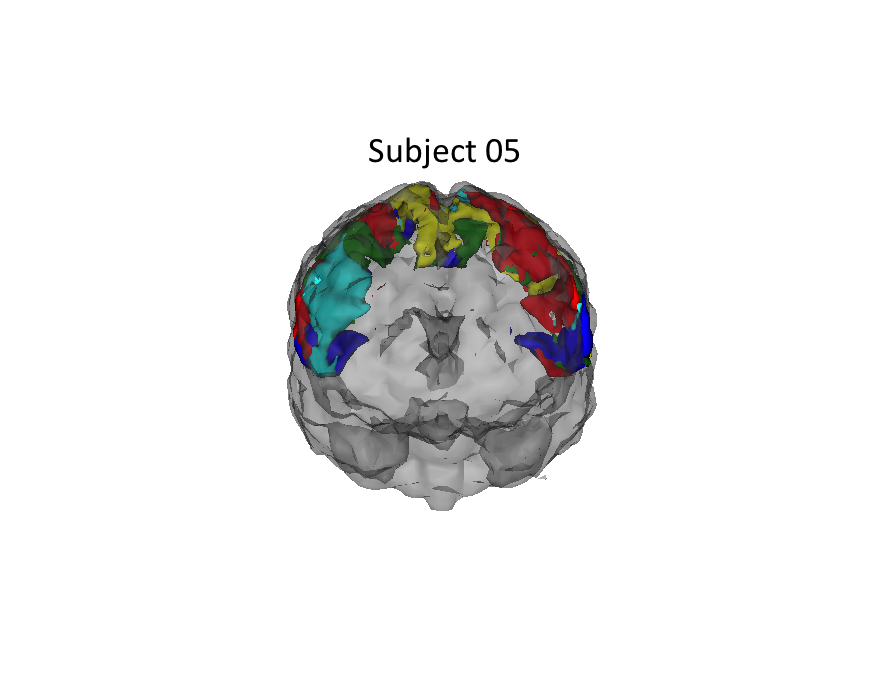}
\caption{Session 1 parcellations - before shrinkage}
\label{fig:parcellations1_before}
\end{subfigure}
\begin{subfigure}[b]{1\textwidth}
\centering
\includegraphics[scale=0.35, trim=9cm 7cm 9cm 6cm, clip]{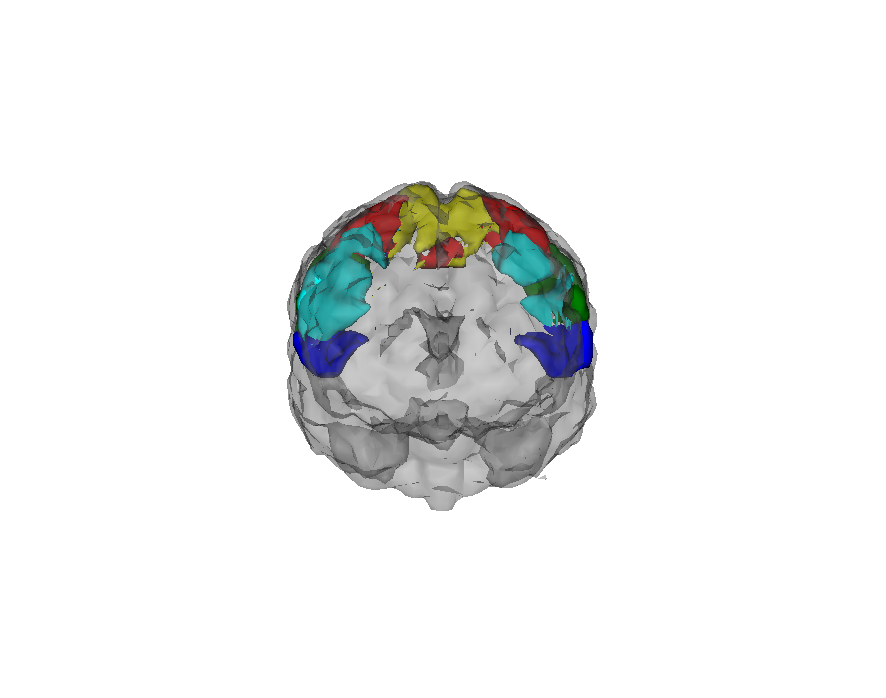}
\includegraphics[scale=0.35, trim=9cm 7cm 9cm 6cm, clip]{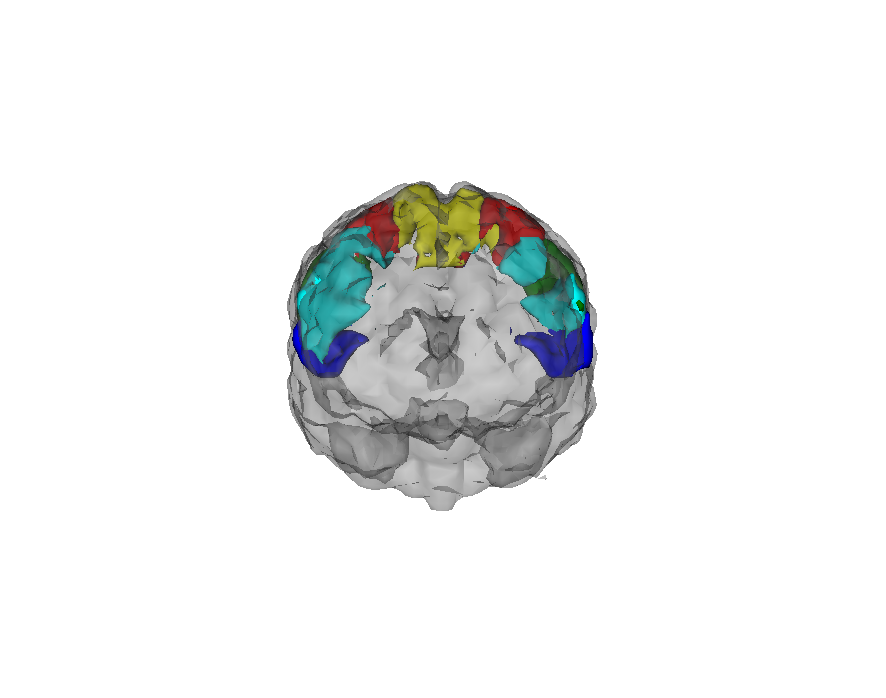}
\includegraphics[scale=0.35, trim=9cm 7cm 9cm 6cm, clip]{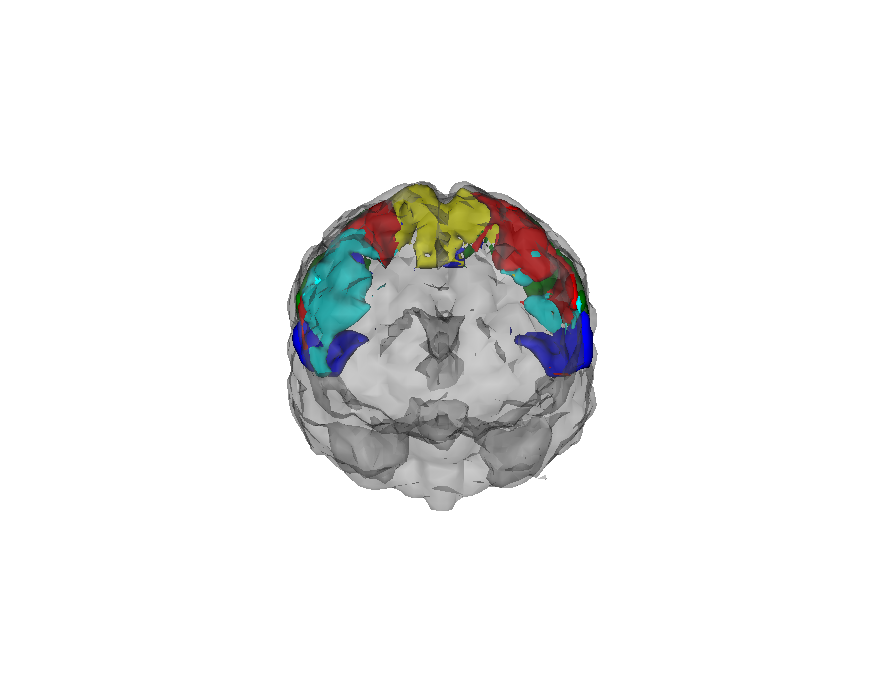}
\caption{Session 1 parcellations - after shrinkage}
\label{fig:parcellations1_after}
\end{subfigure}
\begin{subfigure}[b]{1\textwidth}
\centering
\includegraphics[scale=0.35, trim=9cm 7cm 9cm 6cm, clip]{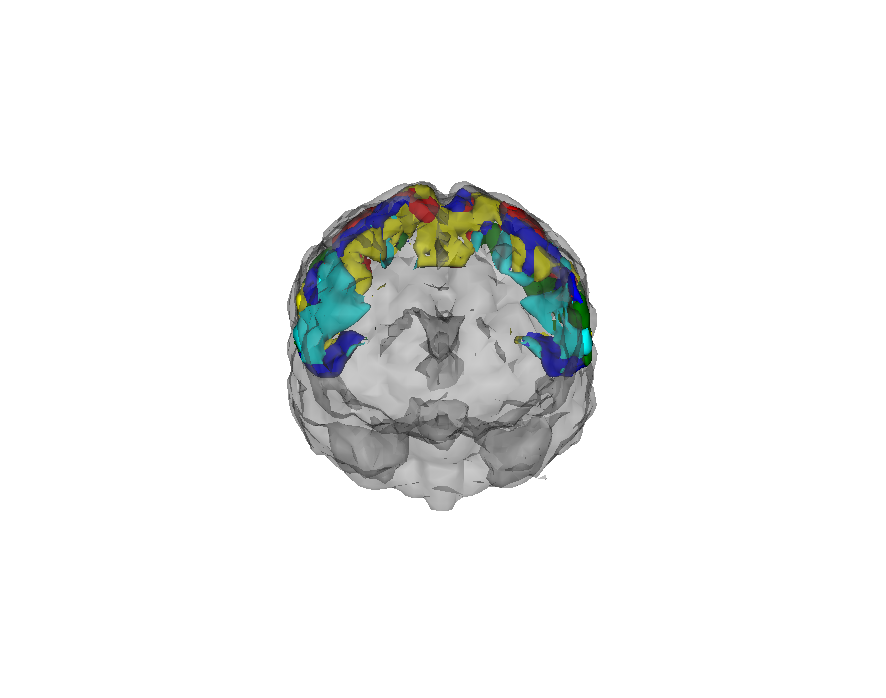}
\includegraphics[scale=0.35, trim=9cm 7cm 9cm 6cm, clip]{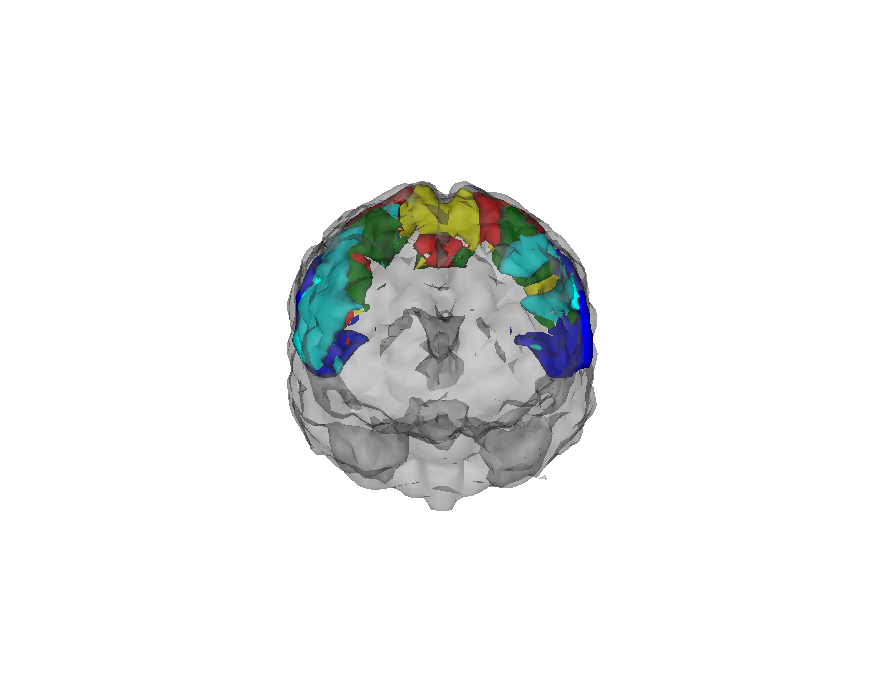}
\includegraphics[scale=0.35, trim=9cm 7cm 9cm 6cm, clip]{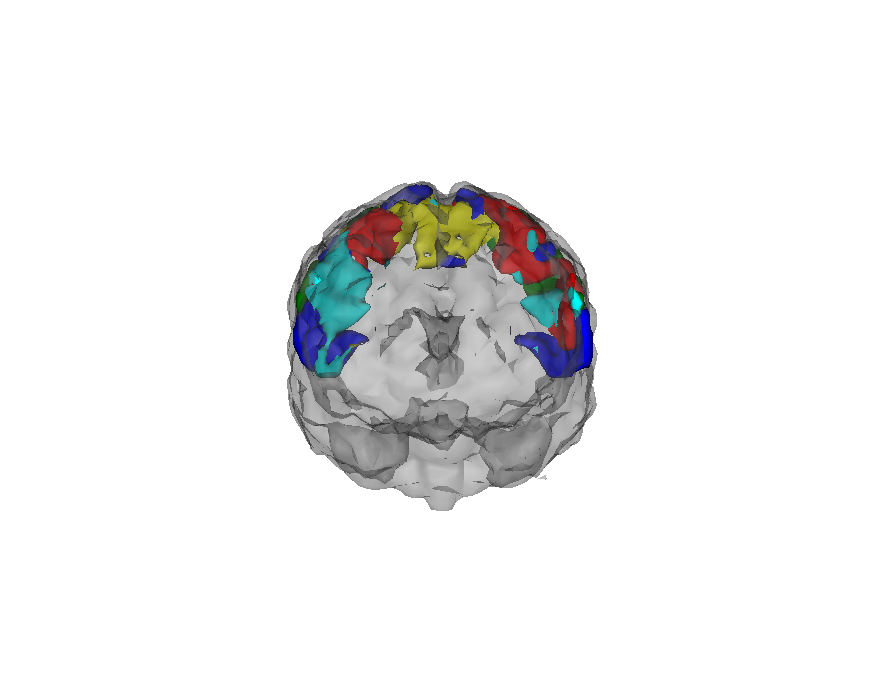}
\caption{Session 2 parcellations - no shrinkage}
\label{fig:parcellations2}
\end{subfigure}
\caption{Subject-level parcellations of the motor cortex from 3 example subjects.\\[28pt]}
\label{fig:parcellations}

\begin{subfigure}[b]{0.5\textwidth}
\centering
\includegraphics[scale=0.4, trim=9cm 7cm 9cm 6cm, clip]{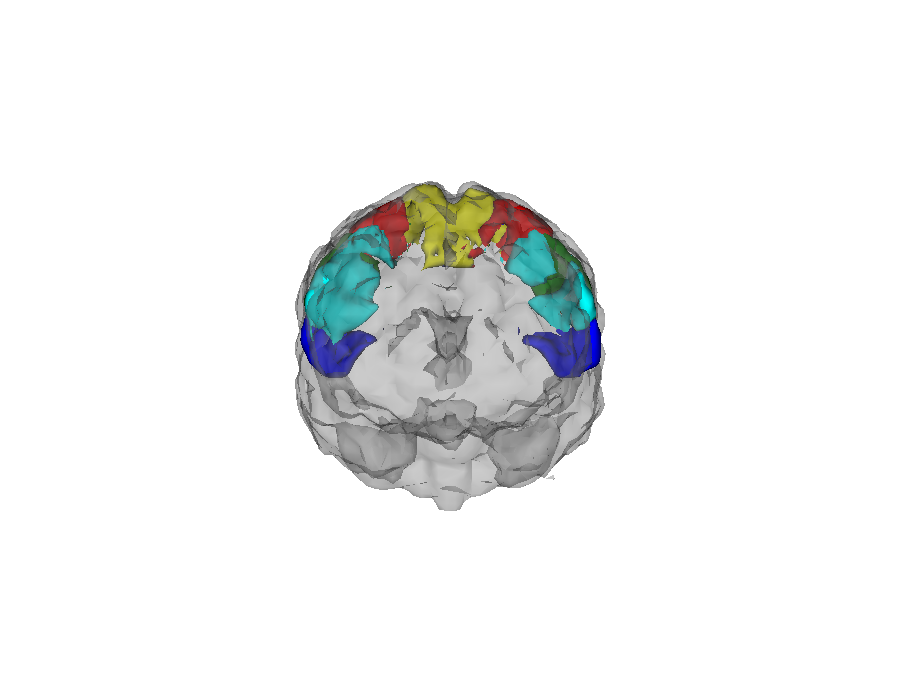}
\caption{Session 1 group-level parcellation}
\label{fig:grp_parcellation1}
\end{subfigure}
\begin{subfigure}[b]{0.5\textwidth}
\centering
\includegraphics[scale=0.4, trim=9cm 7cm 9cm 6cm, clip]{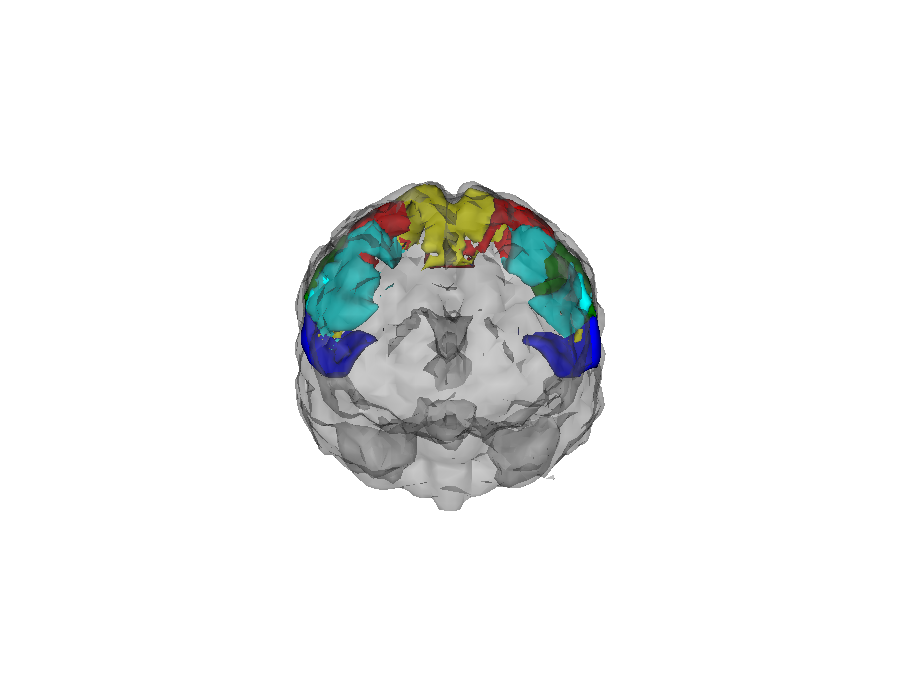}
\caption{Session 2 group-level parcellation}
\label{fig:grp_parcellation2}
\end{subfigure}
\caption{Group-level parcellations from sessions 1 and 2.}
\label{fig:grp_parcellation}
\end{figure}

\textbf{Analysis R2: Performance of parcellations}

Figure \ref{fig:Dice} shows the Dice coefficients of similarity (with test set parcellations) of the parcellations generated from the raw and shrinkage correlation estimates, by noise variance estimation method and type of dataset (single session or test-retest) used to perform shrinkage.  As in Figure \ref{fig:MSE}, each line shows the results for a single subject, and the boxplots show the distributions of Dice coefficient values over all subjects.  For each noise variance estimation method, the percent increase in the median Dice coefficient of the shrinkage parcellations, compared with the raw parcellations, is reported below each boxplot.

Since shrinkage on the correlation estimates without Fisher-transforming resulted in better performance than shrinkage on Fisher-transformed correlations, parcellations were only generated from the shrinkage estimates obtained by shrinking the correlations directly.  The results below therefore reflect the performance of parcellations obtained using this method.

All shrinkage methods resulted in an increase in the median Dice coefficient of parcellations compared with raw parcellations.  When a single session was used to compute the noise variance, the raw parcellations had a median Dice coefficient of 0.401.  The shrinkage-based parcellations had a median Dice coefficient of 0.511 (27.6\% higher) with a common noise variance; 0.521 (30.0\% higher) with individual noise variance; 0.512 (27.8\% higher) with scaled noise variance; and 0.521 (30.0\% higher) with a global noise variance.  When test-retest data was used to compute the noise variance, the raw parcellations had a median Dice coefficient of 0.403.  The shrinkage-based parcellations had a median Dice coefficient of 0.467 (15.9\% higher) with a common noise variance; 0.461 (14.4\% higher) with individual noise variance; 0.465 (15.5\% higher) with scaled noise variance; and 0.479 (18.8\% higher) with a global noise variance.

The improvement in test-retest reliability of parcellations due to shrinkage was remarkably similar across shrinkage methods.  Whether single session data or test-retest data was used to compute the noise variance, the global noise variance estimator again showed the best performance, with an increase in the Dice coefficient of 30.0\% using single session data or 18.8\% using test-retest data.

Figure \ref{fig:parcellations} shows the parcellations of the motor cortex of three subjects (from left to right: subjects 2, 3 and 5) resulting from the first scanning session before shrinkage (Figure \ref{fig:parcellations1_before}), the first scanning session after shrinkage (Figure \ref{fig:parcellations1_after}), and the second scanning session with no shrinkage (Figure \ref{fig:parcellations2}).  The parcellations in Figure \ref{fig:parcellations1_after} were based on shrinkage using a single scanning session and the global noise variance estimator.  These parcellations illustrate that subject-level differences in parcellations generated from raw correlation estimates are not always seen in subsequent scanning sessions.  They also illustrate that while shrinkage-based parcellations are, by nature, more similar to the group-level parcellation, shown in Figure \ref{fig:grp_parcellation}, subject-level differences can still be seen.

\textbf{Time series length and noise variance}

\begin{figure}
\begin{subfigure}[b]{0.5\textwidth}
\centering
\includegraphics[scale=0.5, trim=0 0 25mm 0, clip]{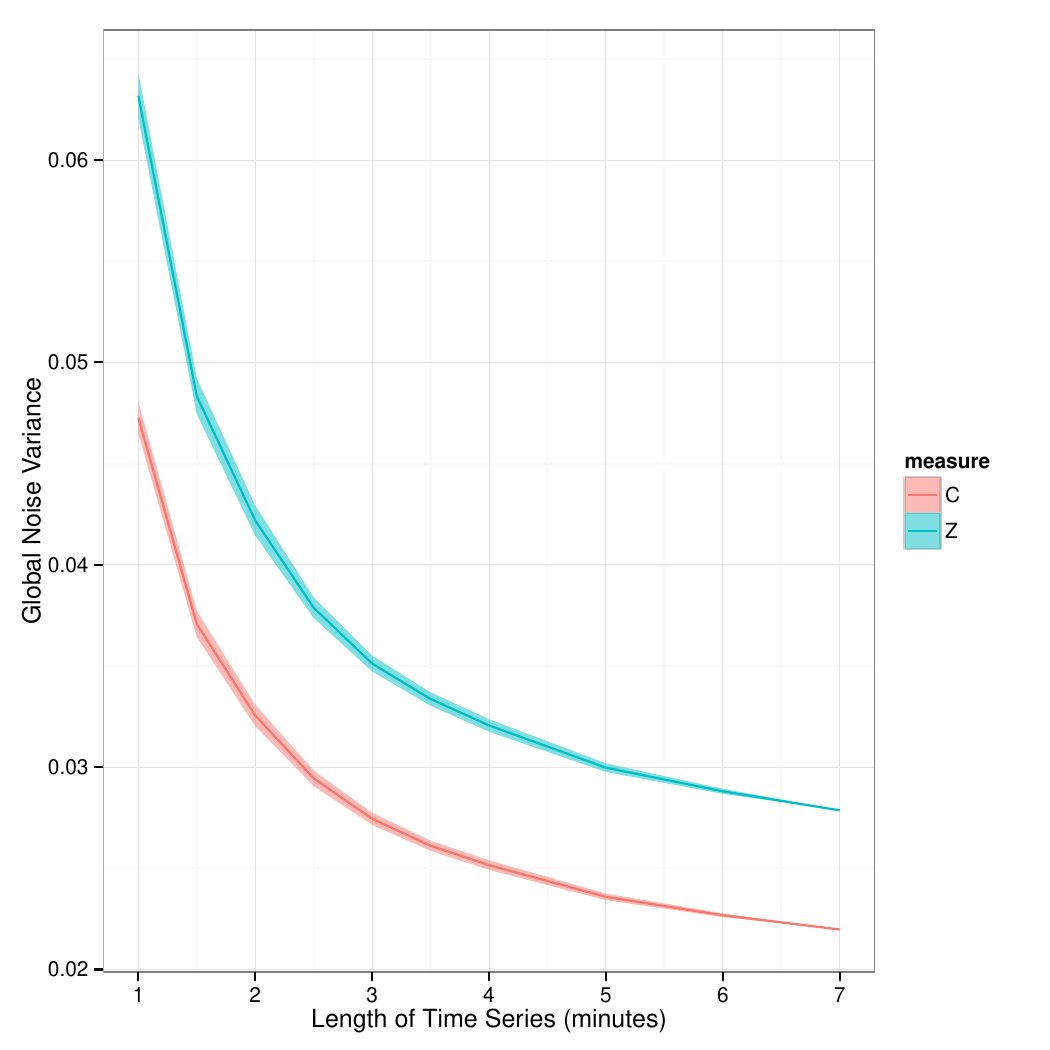}
\caption{Time series length $t$ vs. noise variance $\sigma_U^2(t)$}
\label{fig:length_vs_noisevar}
\end{subfigure}
\begin{subfigure}[b]{0.5\textwidth}
\centering
\includegraphics[scale=0.5]{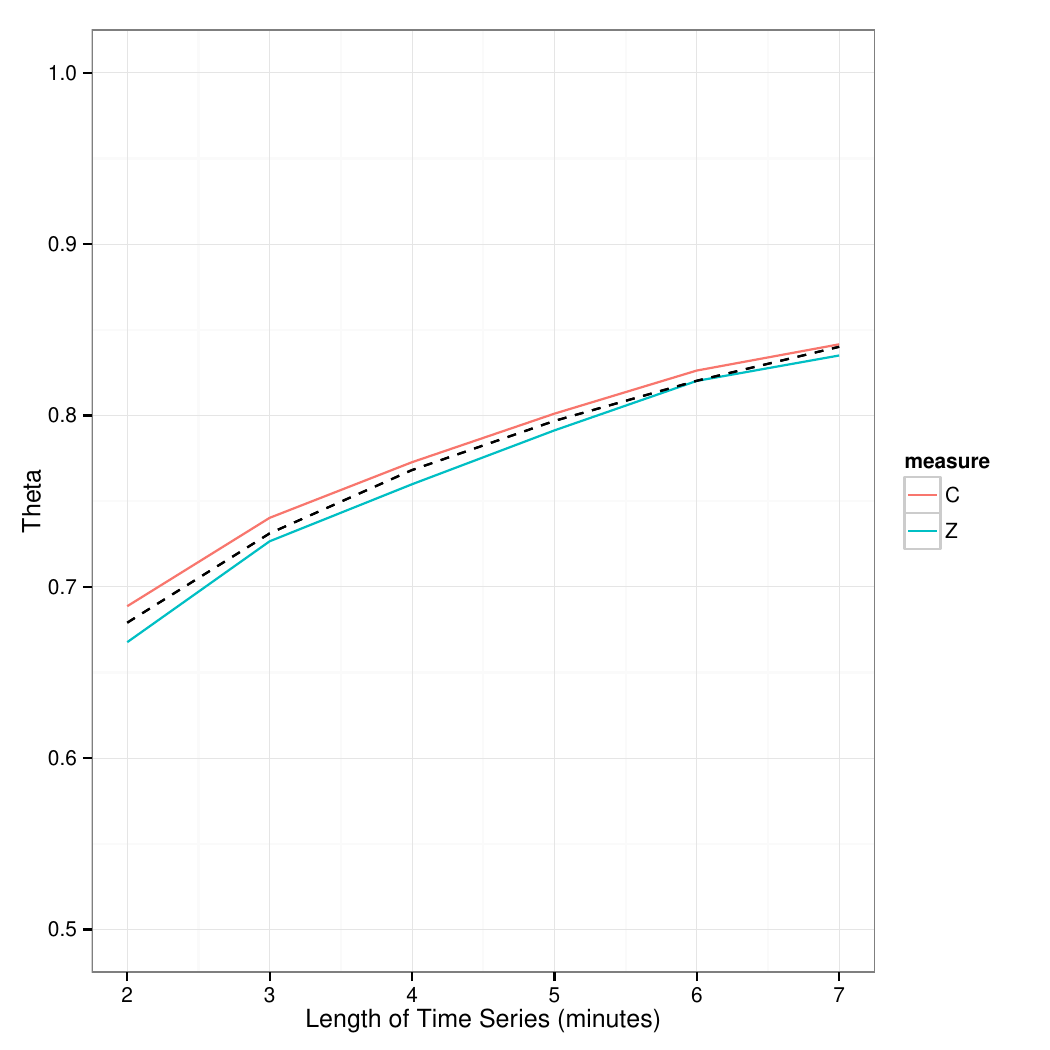}
\caption{Time series length $t$ vs. $\theta(t)=\sigma_U^2(t)/\sigma_U^2(t/2)$}
\label{fig:length_vs_theta}
\end{subfigure}
\caption{The relationship between scan length and noise variance.  Results are shown using untranformed correlations (C, shown in red) and Fisher-transformed correlations (Z, shown in teal) to compute the noise variance.  Panel (a) shows diminishing session-to-session variance as scan length increases.  Panel (b) shows how the adjustment factor $\theta(t)=\sigma_U^2(t)/\sigma_U^2(t/2)$ changes as $t$ increases.  The fitted line relating $\theta(t)$ to $\log(t)$ is shown in black.}
\label{fig:length}
\end{figure}

Figure \ref{fig:length} shows the estimated noise variance $\sigma_U^2(t)$ with 95\% confidence interval (a) and estimated adjustment factor $\theta(t)=\sigma_U^2(t)/\sigma_U^2(t/2)$ (b) for scan lengths ranging from $t=1$ to 7 minutes.  Results are shown using untranformed correlations and Fisher-transformed correlations to compute the noise variance.  Figure \ref{fig:length_vs_theta} also shows the fitted line from the regression relating $\log(t)$ to $\theta(t)$.  The coefficient estimates and standard errors from the regression model
$$
\theta(t)=\beta_0+\beta_1\times\log(t)+\epsilon,
$$
were $\hat\beta_0=0.590$ (s.e. 0.00732) and $\hat\beta_1=0.129$ (s.e. 0.00493).  The adjusted R-squared of the model was 0.986.

\section{Discussion}

In this work we propose a new approach for improving the test-retest reliability of subject-level resting state parcellations based upon the use of shrinkage-based measures of similarity, or distance, between voxels as input to clustering. On 7-minute resting-state scans from 20 healthy adults, parcellations obtained using shrinkage correlation estimates were shown to have up to 30\% improved test-retest reliability compared to those obtained using the raw correlation estimates. Through simulations, similar improvement in reliability were observed for a wide range of sample sizes, time series lengths, signal strengths, and degrees of similarity among subjects. 

Shrinkage methods have found wide usage in the statistics literature, allowing noisy subject-level estimators to ``borrow strength'' from a larger population of subjects. The approach is implicit in penalized likelihood inference, multi-level models (\citealt{lindquist2009correlations}), empirical Bayes estimation (\citealt{friston2002classical}; \citealt{friston2003posterior}; \citealt{su2009modified}), and Bayesian inference. Indeed, most shrinkage estimators will correspond to the mode of some Bayesian posterior. In recent work, Shou et al. (2014) applied shrinkage to rsfMRI seed-based connectivity analysis and showed a nearly 30\% average improvement in intra-subject reliability of correlation estimates. Our work extends these results by developing shrinkage estimators for the full voxel-by-voxel distance matrix required for clustering; proposing methods for constructing shrinkage estimators when only a single scan is available; exploring the utility of shrinkage estimators when the amount of shrinkage is subject-dependent to account for differences in intra-subject variability; and demonstrating improved test-retest reliability of subject-level parcellations based on shrinkage estimates.

Typically, subject-level parcellations derived from short rsfMRI scans (e.g., 5-10 minutes) tend to be highly unreliable due to their low SNR. Longer scans (e.g., 30-60 minutes) lead to more reliable results, and many subject-level parcellation methods are taking advantage of the increased availability of such data.  However, there are still a number of reasons why it may be useful or necessary to use shorter scans to produce subject-level parcellations. First, many such rsfMRI scans have already been collected, from which a wealth of information is potentially available. Second, it may be infeasible to collect longer scans for certain populations, including children, the elderly, or diseased populations.  While healthy adult controls are ideal candidates to undergo long resting-state scanning sessions, they are not always of primary interest to researchers. Third, the price of obtaining long scans may be prohibitive.  By borrowing strength from the group mean to enhance the quality of noisy subject-level functional connectivity estimates, our shrinkage methods are able to minimize the limitations of short rsfMRI scans and take advantage of the widespread availability of such scans for subject-level inference.

Furthermore, our simulations suggest that shrinkage-based parcellations derived from short scans (200 time points) are equivalent, in terms of reliability, to raw parcellations derived from much longer scans (over 1000 time points). This suggests that performing shrinkage may be comparable to collecting scans of considerably longer length, in terms of reliability.  This is an important finding, due to the high costs involved with performing longer scans.  Nonetheless, it may still be beneficial to collect longer rsfMRI scans when feasible. More data is almost always better, and subject-level differences will likely be more accurately expressed as the quantity of subject-level data increases. On the other hand, longer scans should not be viewed as mutually exclusive with shrinkage methods. Our simulation results suggest that shrinkage on longer scans can still lead to substantially more reliable subject-level parcellations. Further research on the benefits of shrinkage on longer scans using real rsfMRI data will be important to understand this interplay.

Finally, according to our simulations, the shrinkage methods we have proposed not only improve the reliability of the overall parcellations, but also the reliability within regions where subject-level differences exist. The ability to accurately parcellate these regions is vital to quantifying and studying subject-level differences in functional organization. 

Although we chose to demonstrate the applicability of our shrinkage methods using correlations as our similarity metric, it is important to note that they are applicable to almost any similarity or distance metric. As long as the assumptions of Normality and signal-noise independence are roughly satisfied, our methods can be applied to a wide variety of metrics. Interestingly, we observed more improvement in reliability by applying shrinkage directly to correlation estimates, rather than to Fisher-transformed correlations, even though the model assumptions are not strictly satisfied for correlation estimates. This is no doubt due to the fact that we are evaluating the error on the correlation scale. If we were instead to evaluate the error on the Fisher-transformed scale, it would be better to perform shrinkage directly on the Fisher-transformed data. We have also applied shrinkage to the inter-voxel similarity metric described in \citeauthor{nebel2012} (\citeyear{nebel2012}), which incorporates long-range correlations between voxels within the motor cortex and the rest of the brain, and found that shrinkage reduced the MSE of the estimates by approximately 40\% (results not shown). Furthermore, though we used spectral clustering for parcellation, there is nothing that prevents other clustering methods from being used instead. For example, we have observed similar improvement in parcellation using K-means clustering (results not shown). The objective of the methods we have described is to maximize the reliability of the similarity or distance metric utilized in clustering. Therefore, any clustering method that depends on such a metric will benefit from using shrinkage estimates of that metric in place of raw observed values.

Throughout the paper we have described a number of potential shrinkage methods. Based on our empirical findings we recommend utilizing the global noise variance estimator, since it is practical for settings where test-retest data is not available (or is only available for a subset of subjects), and it demonstrated the best performance in terms of improved reliability of correlation estimates and parcellations. This method has been implemented in both MATLAB and R and is available for download through Mathworks File Exchange\footnote{http://www.mathworks.com/matlabcentral/fileexchange/48453-mandymejia-shrinkit} and GitHub\footnote{https://github.com/mandymejia/shrinkR}, respectively.  The strong performance of the global noise variance estimator does not necessarily imply that there is no spatial variability in noise levels, but rather that there may be too many parameters to estimate in a meaningful way. For example, in the motor cortex alone there are more than 27 million unique voxel-pairs, and estimating a shrinkage parameter for each can be prohibitive. This may also explain why the scaled noise variance method tends to out-perform the individual noise variance method. Though both noise variance estimators seek to quantify each subject's personal noise level, the scaled noise variance estimator requires the estimation of dramatically fewer parameters. For example, with 7396 voxels and 20 subjects, rather than estimating over $20\times$27,000,000 parameters, the scaled noise variance approach requires approximately 27,000,000$ + 20$ parameters. As more data is collected, more parameters can be reliably estimated. Therefore, with longer scans, the common or scaled noise variance estimators may ultimately begin to out-perform the global noise variance estimator.  However, there is another issue to keep in mind when performing subject-specific shrinkage using the scaled or individual noise variance estimator. When all subjects share the same shrinkage parameter, the rank of the subjects' values relative to one other will not change as each subject's value changes. By contrast, if shrinkage is subject-specific, the rank of subjects may not be preserved. Therefore, even if subject-level parameters can be reliably estimated, care should be taken when performing subject-specific shrinkage.

Although we have strong evidence of the benefits of the proposed shrinkage methods, our analysis was limited to the motor cortex in a population of healthy adults. The benefits of shrinkage on reproducibility are likely to vary, depending on the inter-voxel similarity metric being estimated, the ROI being parcellated, and the population being studied, among other factors. Future research should focus on quantifying the benefits of shrinkage within other regions of interest, for whole-brain parcellation, and for other, potentially more diverse, populations. Finally, while the methods we have proposed can be easily applied to any distance or similarity metric, some parcellation methods employ other versions of subject-level data, such as the entire time series or principal components. Future research should focus on adapting the ideas presented in this paper to such settings.

\section*{Acknowledgements}

This material is based upon work supported by the National Science Foundation Graduate Research Fellowship Program under Grant No. DGE-1232825. This research was supported in part by NIH grants R01 EB016061 and P41 EB015909 from the National Institute of Biomedical Imaging and Bioengineering.  

\section*{Appendix}

\textit{Claim.} The expected value of the common noise variance estimator and the expected value of the mean individual noise variance estimator are the same at each voxel-pair $(v,v')$.
$$
E\left[\hat\sigma_U^{2(C)}(v,v')\right]=
E\left[\frac{1}{I}\sumi \hat\sigma_{U,i}^{2(I)}(v,v')\right]
$$
Starting from the LHS and dropping the $(v,v')$ notation for conciseness, we see that
\begin{align*}
E\left[\hat\sigma_U^{2(C)}\right] 
&= E\left[\frac{1}{2(I-1)}\sumi\left(D_i-\bar{D}\right)^2\right] \\
&= \frac{1}{2(I-1)}E\left[\sumi\left(D_i^2-2D_i\bar{D}+\bar{D}^2\right)\right] \\
&= \frac{1}{2(I-1)}E\left[\sumi D_i^2 -2\bar{D}\sumi D_i+I\bar{D}^2\right] \\
&= \frac{1}{2(I-1)}E\left[\sumi D_i^2 -2I\bar{D}^2+I\bar{D}^2\right] \\
&= \frac{1}{2(I-1)}E\left[\sumi D_i^2 -I\bar{D}^2\right] \\
&= \frac{1}{2(I-1)}\sumi E\left[D_i^2\right] 
    - \frac{I}{2(I-1)}E\left[\bar{D}^2\right] \\
&= \frac{1}{2(I-1)}\sumi E\left[2\hat\sigma_{U,i}^{2(I)}\right] 
    - \frac{I}{2(I-1)}Var(\bar{D}) \\
&= \frac{1}{I-1}\sumi \sigma_{U,i}^2 
    - \frac{I}{2(I-1)}\frac{1}{I^2}\sumi Var\left(D_i\right) \\
&= \frac{1}{I-1}\sumi\sigma_{U,i}^2
    - \frac{1}{2I(I-1)}\sumi Var\left(U_{i2}-U_{i1}\right) \\
&= \frac{1}{I-1}\sumi\sigma_{U,i}^2
    - \frac{1}{2I(I-1)}\sumi 2Var\left(U_{ij}\right) \\
&= \frac{1}{I-1}\sumi\sigma_{U,i}^2
    - \frac{1}{I(I-1)}\sumi\sigma_{U,i}^2 \\
&= \frac{1}{I}\sumi\sigma_{U,i}^2
\end{align*}

Equality with the RHS clearly follows, since
$$
E\left[\frac{1}{I}\sumi \hat\sigma_{U,i}^{2(I)}\right]
=\frac{1}{I}\sumi E\left[\hat\sigma_{U,i}^{2(I)}\right]
=\frac{1}{I}\sumi \sigma_{U,i}^2.
$$

\bibliography{FullPaper}

\begin{thebibliography}{}

\bibitem[Basu et~al., 2002]{basu2002semi}
Basu, S., Banerjee, A., and Mooney, R.~J. (2002).
\newblock Semi-supervised clustering by seeding.
\newblock In {\em ICML}, volume~2, pages 27--34.

\bibitem[Baumgartner et~al., 1997]{baumgartner1997fuzzy}
Baumgartner, R., Scarth, G., Teichtmeister, C., Somorjai, R., and Moser, E.
  (1997).
\newblock Fuzzy clustering of gradient-echo functional MRI in the human visual
  cortex. Part I: Reproducibility.
\newblock {\em Journal of Magnetic Resonance Imaging}, 7(6):1094--1101.

\bibitem[Behzadi et~al., 2007]{behzadi2007component}
Behzadi, Y., Restom, K., Liau, J., and Liu, T.~T. (2007).
\newblock A component based noise correction method (CompCor) for BOLD and
  perfusion based fMRI.
\newblock {\em Neuroimage}, 37(1):90--101.

\bibitem[Bellec et~al., 2010]{bellec2010multi}
Bellec, P., Rosa-Neto, P., Lyttelton, O.~C., Benali, H., and Evans, A.~C.
  (2010).
\newblock Multi-level bootstrap analysis of stable clusters in resting-state
  fMRI.
\newblock {\em Neuroimage}, 51(3):1126--1139.

\bibitem[Blumensath et~al., 2013]{blumensath2013spatially}
Blumensath, T., Jbabdi, S., Glasser, M.~F., Van~Essen, D.~C., Ugurbil, K.,
  Behrens, T.~E., and Smith, S.~M. (2013).
\newblock Spatially constrained hierarchical parcellation of the brain with
  resting-state fMRI.
\newblock {\em Neuroimage}, 76:313--324.

\bibitem[Carroll et~al., 2006]{carroll2006measurement}
Carroll, R.~J., Ruppert, D., Stefanski, L.~A., and Crainiceanu, C.~M. (2006).
\newblock {\em Measurement error in nonlinear models: a modern perspective}.
\newblock Chapman and Hall/CRC.

\bibitem[Cohen et~al., 2008]{cohen2008}
Cohen, A.~L., Fair, D.~A., Dosenbach, N.~U., Miezin, F.~M., Dierker, D.,
  Van~Essen, D.~C., Schlaggar, B.~L., and Petersen, S.~E. (2008).
\newblock Defining functional areas in individual human brains using resting
  functional connectivity MRI.
\newblock {\em Neuroimage}, 41(1):45.

\bibitem[Cordes et~al., 2002]{cordes2002hierarchical}
Cordes, D., Haughton, V., Carew, J.~D., Arfanakis, K., and Maravilla, K.
  (2002).
\newblock Hierarchical clustering to measure connectivity in fMRI resting-state
  data.
\newblock {\em Magnetic resonance imaging}, 20(4):305--317.

\bibitem[Craddock et~al., 2011]{craddock2011}
Craddock, R.~C., James, G.~A., Holtzheimer, P.~E., Hu, X.~P., and Mayberg,
  H.~S. (2011).
\newblock A whole brain fMRI atlas generated via spatially constrained spectral
  clustering.
\newblock {\em Human brain mapping}, 33(8):1914--1928.

\bibitem[Damoiseaux et~al., 2006]{damoiseaux2006}
Damoiseaux, J., Rombouts, S., Barkhof, F., Scheltens, P., Stam, C., Smith,
  S.~M., and Beckmann, C. (2006).
\newblock Consistent resting-state networks across healthy subjects.
\newblock {\em Proceedings of the national academy of sciences},
  103(37):13848--13853.

\bibitem[De~Luca et~al., 2006]{deluca2006}
De~Luca, M., Beckmann, C., De~Stefano, N., Matthews, P., Smith, S.~M., et~al.
  (2006).
\newblock fMRI resting state networks define distinct modes of long-distance
  interactions in the human brain.
\newblock {\em Neuroimage}, 29(4):1359--1367.

\bibitem[Efron and Morris, 1975]{efron1975data}
Efron, B. and Morris, C. (1975).
\newblock Data analysis using Stein's estimator and its generalizations.
\newblock {\em Journal of the American Statistical Association},
  70(350):311--319.

\bibitem[Friston and Penny, 2003]{friston2003posterior}
Friston, K. and Penny, W. (2003).
\newblock Posterior probability maps and SPMs.
\newblock {\em Neuroimage}, 19(3):1240--1249.

\bibitem[Friston et~al., 2002]{friston2002classical}
Friston, K.~J., Penny, W., Phillips, C., Kiebel, S., Hinton, G., and Ashburner,
  J. (2002).
\newblock Classical and Bayesian inference in neuroimaging: theory.
\newblock {\em NeuroImage}, 16(2):465--483.

\bibitem[James and Stein, 1961]{james1961estimation}
James, W. and Stein, C. (1961).
\newblock Estimation with quadratic loss.
\newblock In {\em Proceedings of the fourth Berkeley symposium on mathematical
  statistics and probability}, volume~1, pages 361--379.

\bibitem[Kim et~al., 2010]{kim2010}
Kim, J.-H., Lee, J.-M., Jo, H.~J., Kim, S.~H., Lee, J.~H., Kim, S.~T., Seo,
  S.~W., Cox, R.~W., Na, D.~L., Kim, S.~I., et~al. (2010).
\newblock Defining functional SMA and pre-SMA subregions in human MFC using
  resting state fMRI: functional connectivity-based parcellation method.
\newblock {\em Neuroimage}, 49(3):2375.

\bibitem[Landman et~al., 2011]{landman2011multi}
Landman, B.~A., Huang, A.~J., Gifford, A., Vikram, D.~S., Lim, I. A.~L.,
  Farrell, J.~A., Bogovic, J.~A., Hua, J., Chen, M., Jarso, S., et~al. (2011).
\newblock Multi-parametric neuroimaging reproducibility: A 3-T resource study.
\newblock {\em Neuroimage}, 54(4):2854--2866.

\bibitem[Lindquist and Mejia, 2014]{lindquist2014zen}
Lindquist, M. and Mejia, A. (2015).
\newblock Zen and the art of multiple comparisons.
\newblock {\em Psychosomatic Medicine}.

\bibitem[Lindquist and Gelman, 2009]{lindquist2009correlations}
Lindquist, M.~A. and Gelman, A. (2009).
\newblock Correlations and multiple comparisons in functional imaging: a
  statistical perspective (Commentary on Vul et al., 2009).
\newblock {\em Perspectives on Psychological Science}, 4(3):310--313.

\bibitem[McKeown et~al., 1998]{mckeown1998independent}
McKeown, M.~J., Sejnowski, T.~J., et~al. (1998).
\newblock Independent component analysis of fMRI data: examining the
  assumptions.
\newblock {\em Human brain mapping}, 6(5-6):368--372.

\bibitem[Muschelli et~al., 2014]{muschelli2014reduction}
Muschelli, J., Nebel, M.~B., Caffo, B.~S., Barber, A.~D., Pekar, J.~J., and
  Mostofsky, S.~H. (2014).
\newblock Reduction of motion-related artifacts in resting state fMRI using
  aCompCor.
\newblock {\em NeuroImage}, 96:22--35.

\bibitem[Nebel et~al., 2014]{nebel2012}
Nebel, M.~B., Joel, S.~E., Muschelli, J., Barber, A.~D., Caffo, B.~S., Pekar,
  J.~J., and Mostofsky, S.~H. (2014).
\newblock Disruption of functional organization within the primary motor cortex
  in children with autism.
\newblock {\em Human Brain Mapping}, 35(2).

\bibitem[Ng et~al., 2001]{ng2001spectral}
Ng, A.~Y., Jordan, M.~I., and Weiss, Y. (2001).
\newblock On spectral clustering analysis and an algorithm.
\newblock {\em Proceedings of Advances in Neural Information Processing
  Systems. Cambridge, MA: MIT Press}, 14:849--856.

\bibitem[Oishi et~al., 2009]{oishi2009atlas}
Oishi, K., Faria, A., Jiang, H., Li, X., Akhter, K., Zhang, J., Hsu, J.~T.,
  Miller, M.~I., van Zijl, P., Albert, M., et~al. (2009).
\newblock Atlas-based whole brain white matter analysis using large deformation
  diffeomorphic metric mapping: application to normal elderly and Alzheimer's
  disease participants.
\newblock {\em Neuroimage}, 46(2):486--499.

\bibitem[Penfield and Boldrey, 1937]{penfield1937somatic}
Penfield, W. and Boldrey, E. (1937).
\newblock Somatic motor and sensory representation in the cerebral cortex of
  man as studied by electrical stimulation.
\newblock {\em Brain: A journal of neurology}.

\bibitem[Ryali et~al., 2012]{ryali2012}
Ryali, S., Chen, T., Supekar, K., and Menon, V. (2012).
\newblock A parcellation scheme based on von Mises-Fisher distributions and
  Markov random fields for segmenting brain regions using resting-state fMRI.
\newblock {\em NeuroImage}.

\bibitem[Salvador et~al., 2002]{salvador2002simple}
Salvador, S., Brovelli, A., and Longo, R. (2002).
\newblock A simple and fast technique for on-line fMRI data analysis.
\newblock {\em Magnetic resonance imaging}, 20(2):207--213.

\bibitem[Shou et~al., 2014]{shou2014shrinkage}
Shou, H., Eloyan, A., Nebel, M.~B., Mejia, A., Pekar, J.~J., Mostofsky, S.,
  Caffo, B., Lindquist, M.~A., and Crainiceanu, C.~M. (2014).
\newblock Shrinkage prediction of seed-voxel brain connectivity using resting
  state fMRI.
\newblock {\em NeuroImage}.

\bibitem[Sporns et~al., 2005]{sporns2005human}
Sporns, O., Tononi, G., and K{\"o}tter, R. (2005).
\newblock The human connectome: a structural description of the human brain.
\newblock {\em PLoS computational biology}, 1(4):e42.

\bibitem[Su et~al., 2009]{su2009modified}
Su, S.-C., Caffo, B., Garrett-Mayer, E., and Bassett, S.~S. (2009).
\newblock Modified test statistics by inter-voxel variance shrinkage with an
  application to fMRI.
\newblock {\em Biostatistics}, 10(2):219--227.

\bibitem[Thirion et~al., 2006]{thirion2006dealing}
Thirion, B., Flandin, G., Pinel, P., Roche, A., Ciuciu, P., and Poline, J.-B.
  (2006).
\newblock Dealing with the shortcomings of spatial normalization: Multi-subject
  parcellation of fMRI datasets.
\newblock {\em Human brain mapping}, 27(8):678--693.

\bibitem[Wig et~al., 2013]{wig2013parcellating}
Wig, G.~S., Laumann, T.~O., Cohen, A.~L., Power, J.~D., Nelson, S.~M., Glasser,
  M.~F., Miezin, F.~M., Snyder, A.~Z., Schlaggar, B.~L., and Petersen, S.~E.
  (2013).
\newblock Parcellating an individual subject's cortical and subcortical brain
  structures using snowball sampling of resting-state correlations.
\newblock {\em Cerebral Cortex}, page bht056.

\bibitem[Yeo et~al., 2011]{yeo2011organization}
Yeo, B.~T., Krienen, F.~M., Sepulcre, J., Sabuncu, M.~R., Lashkari, D.,
  Hollinshead, M., Roffman, J.~L., Smoller, J.~W., Z{\"o}llei, L., Polimeni,
  J.~R., et~al. (2011).
\newblock The organization of the human cerebral cortex estimated by intrinsic
  functional connectivity.
\newblock {\em Journal of neurophysiology}, 106(3):1125--1165.

\bibitem[Zilles and Amunts, 2010]{zilles2010centenary}
Zilles, K. and Amunts, K. (2010).
\newblock Centenary of Brodmann's map--conception and fate.
\newblock {\em Nature Reviews Neuroscience}, 11(2):139--145.

\end{thebibliography}
\bibliographystyle{apalike}

\end{document}